\documentclass[journal]{IEEEtran}

\usepackage{graphicx}
\usepackage{latexsym}
\usepackage{amsmath}
\usepackage{hyperref}
\usepackage{pxfonts}
\DeclareGraphicsExtensions{.pdf,.png,.jpg}

\title{Motion of the plasma critical layer during relativistic-electron laser interaction with immobile and comoving ion plasma for ion acceleration}

\author{Aakash A. Sahai  \thanks{aakash.sahai@gmail.com} \\Dept of Electrical Engineering, Duke University, Durham, NC, 27708 USA}

\begin{document}

\maketitle

\begin{abstract}
We analyze the motion of the plasma critical layer by two different processes in the relativistic-electron laser-plasma interaction regime ($a_0>1$). The differences are highlighted when the critical layer ions are stationary in contrast to when they move with it. Controlling the speed of the plasma critical layer in this regime is essential for creating low-$\beta$ traveling acceleration structures of sufficient laser-excited potential for laser ion accelerators (LIA). In Relativistically Induced Transparency Acceleration (RITA) scheme the heavy plasma-ions are fixed and only trace-density light-ions are accelerated. The relativistic critical layer and the acceleration structure move longitudinally forward by laser inducing transparency through apparent relativistic increase in electron mass. In the Radiation Pressure Acceleration (RPA) scheme the whole plasma is longitudinally pushed forward under the action of the laser radiation pressure, possible only when plasma ions co-propagate with the laser front. In RPA the acceleration structure velocity critically depends upon plasma-ion mass in addition to the laser intensity and plasma density. In RITA, mass of the heavy immobile plasma-ions does not affect the speed of the critical layer. Inertia of the bared immobile ions in RITA excites the charge separation potential whereas RPA is not possible when ions are stationary.

\end{abstract}

\small{

\section{Introduction}

\begin{figure}
	\begin{center}
   	\includegraphics[width=3.4in]{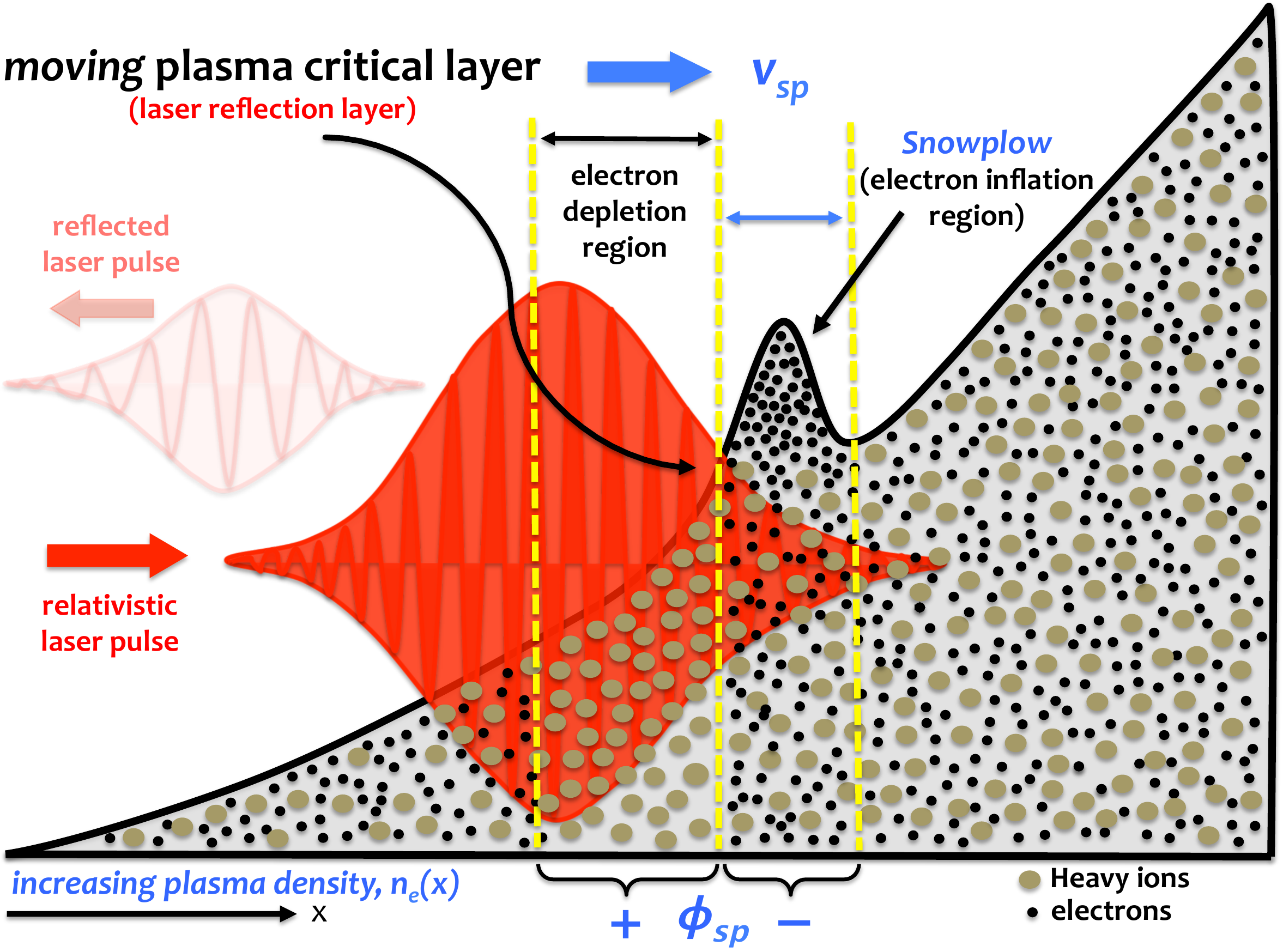}
	\end{center}
\caption{ { \scriptsize {\bf Critical layer motion by relativistically induced transparency in an immobile-ion plasma}. During the interaction of an ultra-short laser with a heavy ion overdense plasma, the plasma ions do not move. While the plasma electrons interacting with the laser-field gain relativistic quiver momentum. Due to coherent relativistic suppression of the collective plasma-electron oscillation frequency the laser pulse can propagate beyond the initial critical layer. The plasma density where the relativistic plasma frequency equals the laser frequency is at a higher electron density in the longitudinally forward direction. The laser ponderomotive force creates a snowplow of electrons, the acceleration structure. By controlling the speed of propagation and potential of the snowplow, trace light-ions are accelerated. The speed control also limits the impulse of the snowplow potential on the heavy plasma-ions.}}
\label{fig:heavy-ion-plasma-concept}
\end{figure}

In this paper we analyze the longitudinal motion of plasma critical layer in the relativistic-electron laser interaction regime. The primary objective of this work is to differentiate the two different physical processes of laser-plasma interaction responsible for the critical layer motion, relativistically induced transparency and radiation pressure. We study these phenomena by contrasting the dynamics in two different types of plasmas. The plasma critical layer is defined as the density contour where the laser stops and reflects. The overdense plasma densities higher than the critical density are defined with respect to the laser frequency.  

It is shown that in a plasma with stationary ions, laser front reflects off the relativistic critical layer. During this process the laser reflection layer moves in the direction of laser propagation as long as the laser amplitude increases above the relativistic intensity. Whereas even a temporally flat-top laser pulse reflected at the critical layer results in a force that causes bulk plasma propagation. This force originates due to momentum conservation resulting from the reversal of the photon momentum flux of the reflected laser. It is shown that speed of the radiation pressure driven laser front is controlled by the plasma ion mass. The classification of ions as light depends on their co-propagation resulting in a bulk plasma radiation pressure front with significant longitudinal speed. In a plasma with infinitely massive plasma-ions, the whole plasma cannot move with the laser front. As will be shown, relativistic effect induced transparency allows laser propagation into an infinite ion-mass plasma. The speed of the laser front moving by the process of relativistically induced transparency originating in the coherent relativistic plasma-electron dynamics in laser field is shown to be independent of the plasma-ion mass. If the speed of the laser front is controlled then the momentum transferred to the heavy background plasma-ions is limited. We present multi-dimensional particle-in-cell (PIC) simulations to demonstrate the key differences in the physical processes resulting in critical layer motion in the direction of laser propagation.

The significance and the need for studying the motion of the plasma critical layer when interacting with strong fields is analyzed from the first principles with application to laser ion acceleration (LIA). With the currently foreseen table-top laser technology the relativistic electron momentum regime is most accessible. In the future, if relativistic-ion and non-linear quantum-electrodynamics lasers can be setup on a table-top, these might be more viable technologies for accelerating ions. It is still likely that the relativistic electron laser will be a cheaper and a more compact option. Thus the physics of the relativistic electron regime of lasers is of long-term interest. When high-intensity lasers are incident upon a target it ablates in their pre-pulse and forms a pre-plasma, so the effects at the critical layer in a density gradient are important. The technology of accelerating ions using laser-plasma ion accelerators has a wide variety of applications such as cancer hadron-therapy, radiography, injectors in conventional particle accelerators, high-energy density science etc.

In the following sections using the pertinent background laser-plasma theory and comparisons to other propagating density structures in the plasma, we motivate the study of the plasma critical layer motion. Laser-plasma ion accelerators have the potential to produce ion beams with unprecedented characteristics of ultra-short bunch lengths, 100s of fs and high bunch-charge, $10^{10}$ particles over acceleration length of about 100 microns. The accelerated ions are emitted from an aperture size of the order of the laser pulse radial focal spot. High beam charge in ultra-short bunch lengths are attainable because the acceleration gradients are of the order of the laser field. Thereby the space-charge limitation of conventional injectors is overcome due to the large momentum gain over short duration leading to rapid adiabatic damping (non conservation of the phase-space volume) of accelerated ions at the source. The motion of the plasma critical layer, by relativistic effects or by radiation pressure of the reflected radiation, is applied to creating low-$\beta$ traveling acceleration structures for trapping and accelerating ions. It is crucial to control the speed of the overdense plasma acceleration structures to trap the ions which requires that the kinetic energy of accelerated ion beam is smaller than the laser excited acceleration structure potential in the plasma. 

In a heavy immobile-ion plasma, laser propagates beyond the cold-electron critical layer upon interaction with relativistic intensity laser pulses while the background plasma-ions are stationary. Relativistic intensity and its increasing amplitude is a necessary condition for the motion. The cold-electron critical layer can no more reflect the relativistic laser field, as a result the critical layer moves to its new relativistically corrected density location\cite{Kaw-Dawson}\cite{Akhiezer-Polovin}. The laser imparts relativistic quiver momentum ($\frac{\vec{p}_{\perp}}{m_ec}=\vec{a}(r,t)$) to all interacting plasma electrons coherently due to the first-order Lorentz force from the laser field. The relativistic momentum makes the plasma electrons heavier by the Lorentz factor resulting in the reduction of collective plasma electron oscillation frequency (depends on the local quasi-neutral plasma electron density). The laser then propagates forward until it is shielded at a higher density (a corresponding higher plasma frequency). At each different intensity in the laser pulse envelope there is an intensity dependent relativistically corrected critical layer. This phenomenon of relativistically induced transparency by the laser field while the plasma-ions are fixed is the basis for Relativistically Induced Transparency Acceleration (RITA)\cite{RITA-PRE-2013}. 

The rate at which the transparency is induced is controlled by shape of the laser envelope rise and spatial-rate of increase of plasma density in the laser propagation direction. The speed of the laser front ($v_{sp}$) is controlled by the rate of induced transparency. In the rest-frame of the laser at a relativistically corrected critical layer, the ponderomotive force excites an electron density inflation, a snowplow. This acceleration structure co-propagates with the laser front. The snowplow made of ponderomotive electrons is a space-charge separation from the background plasma-ions which excites an electrostatic field. The field lines in the snowplow are directed from the stationary plasma-ions towards the electrons. The speed of this acceleration structure is controlled by the choice of laser-plasma interaction parameters which control the induced transparency. This tuning of the speed provides control over the beam energy of the doped trace-density light-ions reflected off the snowplow potential. Importantly the control also ensures that the particles to be accelerated are trapped and accelerated to the desired energy by beam-loading the acceleration potential, while also maintaining mono-energeticity. The control over the speed also limits the impulse kick on the heavy plasma-ions due to the snowplow field. We present motion of the laser pulse over many plasma wavelengths and ignore the fast oscillations of the critical layer observed during a single cycle\cite{fast-osc-2004}.

In a light co-moving ion plasma the laser pulse acts as a piston. Relativistic intensity is not a necessary condition. An entire region of the plasma, electrons and ions, irradiated by the laser have to move as one. The motion occurs as all the plasma electrons directly interacting with the laser at the critical density reflect it (with partial absorption). The irradiated plasma electrons at the critical density are then collectively driven forward as an inflation layer by the laser radiation pressure and drag along all the plasma ions. In the RPA schemes\cite{hole-boring-1992}\cite{rpa-2004} the acceleration structure is a double layer of electron and ion inflation layer. The electron layer is driven by the pressure of the laser photon momentum flux reversal and the ion layer is dragged behind the laser front by the electrostatic pull of the electron layer. As the whole plasma is pushed forward at the laser reflection layer, the critical layer moves with the plasma. Therefore the acceleration structure speed in RPA ($v_{rpa}$) depends upon the plasma-ion mass. This is because the radiation pressure driven electron layer cannot move further due to the drag of the plasma ions. The displacement between the electron layer and the plasma ions creates an electrostatic potential. To enable further motion of the critical layer the electrostatic potential of the electron layer has to create an impulse, $F\Delta t$ high enough to pull the plasma ions to co-propagate. This requirement creates a threshold for the laser intensity and pulse duration to excite radiation pressure driven critical layer motion. Since the momentum gained by the plasma-ions depends upon both the magnitude of force and the interaction time, the pulse length plays a significant role. 


\begin{figure}
	\begin{center}
   	\includegraphics[width=3.4in]{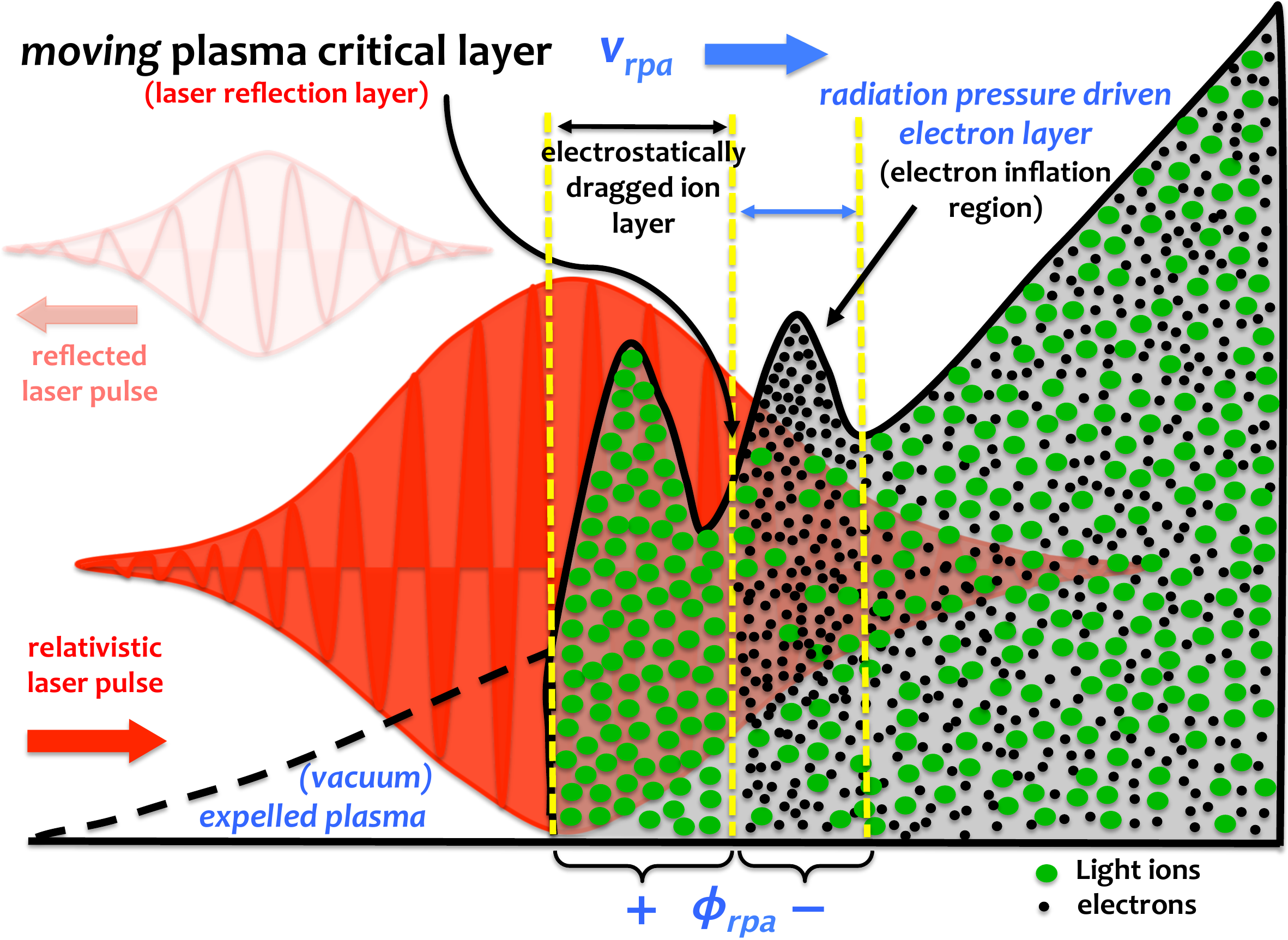}
	\end{center}
\caption{ { \scriptsize {\bf Critical layer motion due to laser radiation pressure dragging the whole plasma}. In a light co-moving ion plasma, the ions are electrostatically pulled forward by the laser-driven electron density layer at the critical surface. Laser reflects off the plasma critical surface and the radiation pressure of photon momentum reversal leads to bulk plasma forward motion. The electrons being lighter move forward with a higher velocity but cannot overcome the electrostatic force of the plasma ions. The plasma ions then move due to a sustained pull from the spatially separated electron layer being continuously driven by the laser radiation pressure. Thus a double layer acceleration structure of laser driven electrons and plasma ions is excited. Its speed is controlled by the mass of the plasma ions in addition to the radiation pressure magnitude and the plasma density.} }
\label{fig:light-ion-plasma-concept}
\end{figure}

When the laser pulse ends, the radiation pressure of the laser stops exerting a forward longitudinal force on the plasma electrons in the focal-spot area. However, at the instant the laser pulse ends there is a large residual longitudinal electron density displacement due to the conservation of longitudinal momentum of photons of the laser stopping and reflecting at the plasma critical layer. After the laser switches off, a double layer forms if the residual plasma electron longitudinal momentum excites a potential large enough to drag along the light plasma-ions. If the laser imparted electron longitudinal momentum is high enough to accelerate the double layer velocity ($v_{CESA}$) to exceed the plasma sound velocity ($c_s=\sqrt{{T_e}/{M_i}}$, ion vibrational velocity) it results in supersonic (${v_{CESA}}/{c_s}>1$, Mach number) bulk plasma propagation of the whole irradiated region. In any medium the sound energy propagates vibrationally with no bulk medium propagation. Thus a freely propagating double layer of electrons followed by light plasma-ions is formed. This acceleration structure propagates forward without being continuously forced by the laser radiation pressure and thereby it is not inflicted by laser seeded instabilities. Shocks are such freely propagating supersonic plasma density structures. The supersonic density inflation on the driven side is not communicated across the shock. This results in the piling up of plasma density on the driven side against the cold plasma ahead of the shock. The shock is formed by the exchange of momentum from the longitudinally driven electron layer to the plasma ions. This transfer of momentum is mediated by the charge separation electrostatic fields, hence the process of shock formation is collision-less. The collision-less electrostatic shock (CESA)\cite{cesa-1992}\cite{cesa-1997} does not exchange energy with the surrounding plasma through fluid processes (like density waves) because the shock structure formation mechanism is itself inherently supersonic.

The ion density perturbations in the plasma vibrationally propagate (no bulk motion) at the speed of sound. Thereby a cold plasma has a smaller speed of sound, so it is easier to launch a shock. However, collisional pre-heating of the plasma can provide an independent control over the Mach number. The mechanism of energy dissipation of the shock is through the energy transfer to light-ions in the cold plasma ahead of the shock front. The energy is transferred to the plasma ions ahead of the shock due to their reflection off the double layer electrostatic field. Thereby CESA is also a process where the plasma ions have to move to constitute a shock. So the critical layer moves with the whole plasma. Though there is no laser pulse reflecting off the critical density within the shock, any subsequent laser pulses would catch up and get reflected off the propagating shock.

In RITA the heavy plasma ions provide an inertial background for exciting the space-charge electrostatic field much in the same way as in laser electron accelerators. When the transparency is induced at a high rate and the laser front moves fast, then the impulse of the electrostatic field on the heavy plasma-ions is small. The interaction time of the ponderomotively excited electron layer with the heavy plasma-ions is controlled so the plasma-ions at the laser front pickup negligible momentum. Therefore the heavy plasma-ions are assumed to be immobile over the timescales of motion of the laser front and the RITA snowplow. The speed of the laser front is therefore not related to the mass or momentum of the heavy plasma-ions. So the plasma-ion mass independence of $v_{sp}$ implies that $v_{sp}>v_{rpa}$ and $v_{sp}>v_{CESA}$ is generally true. The classification of ions as light or heavy depends upon the mass-to-charge ($\frac{M_{ion}}{Z}$) ratio of the plasma ions. The ionization state of the plasma ions for high-Z atoms in relativistic-electron intensity laser is not well characterized.

In section \ref{section:laser-plasma model} we review the critical layer laser-plasma interaction model relevant to our analysis in the later sections. As a basis for the later sections we briefly introduce the plasma electron motion in relativistic laser field, the speed of the excited plasma space-charge separation fields, their magnitude and their trapping of the plasma particles. In section \ref{section:plasma acc-structure} we put the motion of the plasma critical layer into perspective by relating it to other propagating plasma acceleration structures.  In section \ref{sec:crit-layer motion analysis}, \ref{sec:crit-layer-motion-1d-sim}, \ref{sec:crit-layer-motion-2d-sim} we present analysis and simulations to demonstrate the differences in the critical layer motion. A direct parametric comparison is not possible from these simulations since we present two different physical mechanisms. In section \ref{sec:high intensity 2d-sim} we compare the motion of the plasma critical layer upon interaction with an ultra-intense ($I_0=1.1\times 10^{22}\frac{W}{cm^2}$) ultra-short laser pulse ($30fs$) for a plasma target with light and heavy plasma ions. 


\section{Modeling critical layer laser-plasma interaction}
\label{section:laser-plasma model}

{\it Laser field in plasma} - The relativistic laser-plasma interaction regime is classified based upon the peak intensity that is attained in the focal spot of the laser. When the laser intensity (and field) exceeds a certain threshold, the laser electric field amplitude gets high enough to directly excite the interacting electrons (in plasma or free-space) to relativistic momentum. The laser field is characterized by the {\it laser strength parameter} $a_0$. It is the peak amplitude of the normalized vector potential ($\vec{B}(r,t)=\vec{\nabla}\times\vec{A}(r,t)$) of the laser field $\vec{a}(x,r,t)$. The normalization quantifies the momentum of a laser excited electron $\vec{p}_e$ in vacuum due to the conservation of the canonical momentum $\vec{P}$. When vorticity $\vec{\Lambda}=\nabla\times\vec{P}$ has no temporal variation, then in a plane-wave $\partial_t\vec{P}_e = 0\Rightarrow \partial_t\left(\vec{p}_e-\frac{e\vec{A}}{c}\right)=0$. It is thereby desirable to quantify the laser intensity with $\vec{a}=\frac{\vec{p}_e}{m_ec}=\frac{1}{m_ec}\frac{e\vec{A}}{c}$. The laser electric field amplitude in terms of $\vec{a}(x,r,t)$, its monochromatic sinusoidal evolution in time (with angular frequency $\omega_0$) and a radial focal spatial profile, is $\vec{E}_{laser}(r,t)=-\frac{1}{c}\frac{\partial\vec{A}(r,t)}{\partial t}=\frac{\omega_0}{c}\vec{A}(r,t)=\omega_0\frac{m_ec}{e}\vec{a}(r,t)$. The laser pulse is not a plane wave and thereby $a_0$ characterizes the peak intensity and peak power $P_0 \propto I_0 \propto a_0^2$. In the relativistic regime ($a_0\ge1$), laser excited electron quivers (transverse) at a peak momentum $p^{max}_{\perp}=max(a)=a_0$, such that its kinetic energy equals or exceeds its rest mass. The free-space electron oscillating transversely at $\omega_0$ excited first-order in $\vec{a}$ by a plane-wave transverse electromagnetic mode does not experience first-order longitudinal momentum gain. The dominant mode of a laser excited plasma electron is at $\omega_{pe}$ in a plasma with fixed-ions and its exact dynamics is in \cite{Akhiezer-Polovin}. 

{\it Laser longitudinal force} - Laser-field imparts longitudinal excitation (in the direction of energy flow) to electrons with an initial transverse velocity component. If the laser-electron interaction is not infinite or there is a time-varying field amplitude then the magnetic part of the Lorentz force $\frac{\vec{p}_{\perp}}{m_e\gamma_e}\times \vec{B}$ ($\vec{p}_{\perp}$ is in phase with $\vec{B}$ and $\pi/2$ out of phase with $\vec{E}$ ) causes a net longitudinal force which is second-order in $\vec{a}$. The longitudinal ponderomotive force\cite{ponderomotive-force} is $F^{~ponde}_z \simeq m_e\frac{dp^{\parallel}_e}{dt}\simeq-e\frac{\vec{p}_{\perp}}{\gamma_e}\times \vec{B} =-\frac{m_ec^2}{2\gamma_e} \nabla_z |a|^2$. Therefore, $p^{\parallel}_e \propto |a|^2$. Because this force is a second order ($\propto a^2$) force its characteristic frequency is $2\omega_0$ and has an average term. When $a<1, |a|^2\ll 1$ however in the relativistic regime $a>1, |a|^2 > 1$. In plasma the net longitudinal force is driven by laser ponderomotive force but it evolves during the interaction due to many effects such as space-charge force, laser pulse envelope changes by self-focussing causing effective lensing or density inhomogeneities. Thus ponderomotively exciting propagating electrostatic space-charge structures in the plasma is an efficient way of transforming laser electric field for beam energy gain. This is especially practical considering that the relativistic intensity lasers ionize any material into a plasma. Unlike the material breakdown limit of the accelerating fields in conventional accelerators, plasma cannot be broken (though there are parasitic effects such as self-trapping of plasma electrons and higher ionization of plasma-ions with higher magnitude plasma fields). 

{\it Acceleration structure velocity} - The laser-plasma electron accelerators are high-$\beta$ acceleration structures which use high-$v_{\phi}$ plasma waves. The high velocity acceleration structures of a laser-plasma electron accelerator need sufficiently high space-charge potentials to be excited in the plasma to trap and accelerate relativistic electrons. The laser technology to excite sufficient ponderomotive potentials for electron accelerators delayed them from 1979 \cite{laser-driven} to 2004 \cite{laser-driven-expt}. The coupling of electromagnetic mode to a plasma-electron mode in an unmagnetized plasma is governed by the dispersion relation of the electromagnetic energy exciting a plasma electron density wave\cite{Akhiezer-Polovin}. For plane-wave electromagnetic field at $\omega_0$ interacting with a homogeneous unmagnetized plasma of characteristic frequency $\omega_{pe}$, the dispersion is governed by $\omega_0^2 = \omega_{pe}^2 + c^2k^2$. The laser electric field has a radial 2-D spatial variation and this simple 1-D dispersion characteristic equation does not describe the dispersion accurately. This dispersion relation is for the electromagnetic field exciting only an plasma electron wave and ignores the energy coupling to plasma ions and the corresponding ion-mode dispersion relations. When $\omega_0\gg\omega_{pe}$, the group velocity of the laser pulse (and hence the speed of the laser excited acceleration structure) in the plasma is very close to the speed of light in vacuum, $v_g=c\sqrt{1-\omega_{pe}^2/\omega_0^2} \lesssim c$. The phase velocity of the electron plasma wave is equal to the group velocity of the laser pulse $\beta^{\phi}_{pe}=v^g_{laser}/c$. This implies $\gamma_{pe}=\omega_{0}/\omega_{pe}$. However, if the plasma frequency is nearly equal to the laser frequency $\omega_{pe}\simeq\omega_0$, the group velocity is very low, $v_g\ll c$ (low $\beta$) and depends upon the ratio of the plasma frequency to the laser frequency $\frac{\omega_{pe}}{\omega_0}$. Laser energy propagation is possible only if $\omega_0>\omega_{pe}$. Alternatively, the whole plasma is driven by radiation pressure, this is possible when the plasma-ions move. Because of the fixed ion assumption in the above analysis, under significant plasma-ion motion it is not valid. In mobile-ion plasma the dispersion of the electromagnetic wave in the plasma has to include the plasma-ion motion. As in \cite{Akhiezer-Polovin} when the plasma is cold, the dynamics of the plasma is described by the first-moment $n_e(x,t)$ and full description using the phase-space distribution $f(x,p,t)$) is not necessary. The properties of collective coherent particle oscillations in such a plasma are characterized by the density distribution. Plasma waves and other coherent structures in the plasma do not constitute thermal plasma. When an excited plasma equilibrates into randomness with maximum entropy (stochastically characterized) through coherent-mode mixing then it is considered thermal. The long term energy dissipation of the plasma fields and deconstruction of the density structures is through instabilities, collisions, kinetic effects such as Landau damping etc.

{\it Plasma fields} - The space-charge electric fields that is excited in a plasma depend upon the plasma density, in addition to the magnitude of charge displacement force. It is the charge displacement from equilibrium between the plasma electrons and ions which creates the fields. Therefore larger the electron density (assuming high enough force and energy driving the perturbation) higher the number of electrons that are displaced from equilibrium. Again, single-mode electron plasma frequency ignores any plasma-ion dynamics ($\omega_{pe}^2(x,t)=\frac{4\pi n_e(x,t) e^2}{m_e}$) and is determined only by the plasma electron density, $n_e(x,t)$ (directly from Gauss' law and Newton's second law of motion). If the force acting on the plasma electrons is below the electrostatic force of the background plasma-ions, they oscillate within the electrostatic potential well of the plasma ions. External energy coupled into the plasma excites motion of the electrons (electron gas neutralized by ions) to the leading order at the plasma electron frequency $\omega_{pe}$. The space-charge force equation on a representative laser excited plasma electron longitudinally oscillating ($A_{\parallel}\simeq 0$) in the space charge Langmuir or plasma-electron wave is $d\vec{p}_{e}/dt = -e\vec{E}_{pe}=e\vec{\nabla}\phi_{pm}$. Where $\phi_{pm}$ is the acceleration structure potential excited in the electron plasma wave (in the rest-frame of the wave). This potential is also referred to as ponderomotive potential because it is excited by the ponderomotive force of a laser pulse envelope. In a beam excited plasma the beam current is the driver term of the plasma wave equation \cite{beam-driven}. For a fixed beam energy (mean velocity of beam particles) it is the beam density which represents the driver strength. We can infer the single-electron plasma space-charge field, longitudinally oscillating at $\omega_{pe}$, based upon the electron momentum (longitudinal momentum $~\vec{\beta}_{pm}\gamma_{pm}$) in the rest-frame of the plasma wave, $|\vec{E}_{pe}|=\left(\frac{m_ec\omega_{pe}}{e}\right)\gamma_{pm}\beta_{pm}$. With the relativistic correction (in the rest-frame of the propagating plasma-wave) to the electron plasma frequency, $|\vec{E}_{pe}|=\left(\frac{m_ec\omega_{pe}^{\gamma_e=1}}{e}\right)\sqrt{\frac{\gamma_{pm}^2-1}{\gamma_e}}$. The maximum amplitude of the electron plasma-wave is therefore dependent upon the magnitude of the ponderomotive force. The limit of the maximum longitudinal field while the plasma electrons are still bound to the background ions is referred to as the wave-breaking limit. As the plasma field approaches the wave-breaking limit the ponderomotively driven plasma-electron density is bunched into a tighter volume in the wave-crest leading to the limit of density steepening. Beyond this limit the plasma electrons are not just forming the density structures but as they escape the plasma-ion electrostatic field, they are trapped in the propagating plasma waves and get accelerated. 

{\it Plasma self-trapping} - Plasma electrons creating the plasma electron density wave are trapped in the potential $\phi_{pm}$ of the wave and get accelerated if $|e\phi_{pm}|>(\gamma^{trap}_e-1)m_ec^2$\cite{esarey-pilloff}. Where $|\gamma^{trap}_e\beta^{trap}_e|$ is the trapped electron momentum in the rest-frame of the laser, relative to the plasma wave momentum. The potential $\phi_{pm}$ is in the rest-frame of the laser or the frame co-moving with the plasma-wave. The plasma electrons are oscillating in the plasma wave and thereby the phase of the plasma-wave relative to the plasma electrons decides the magnitude of the relative momentum of the electrons. The laser is at the head of the electron plasma wave (the wave is in the wake of the laser) of relativistic phase velocity which equals the group velocity of the laser. The electron density trapped in the plasma wave is subject to the total energy equation depending on the number density of the ponderomotive electrons. The trapping of plasma electrons and the transfer of energy from the space-charge plasma fields leads to effective lowering of the fields. This is referred to as beam-loading. Similarly, in laser-plasma ion accelerators the transfer of energy from the acceleration structure to the trapped ions is a mechanism of energy loss through beam-loading. 

{\it Plasma fields for LIA} - The maximum electric field of a plasma wave in the non-relativistic excitation of plasma electrons is $|\vec{E}_{pe}|=\omega_{pe}\left(\frac{m_ec}{e}\right)$. Therefore in the non-relativistic regime the limit of the plasma electric field is $E_{pe} \propto \omega_{pe} \propto \sqrt{n_e}$. So the maximum longitudinal electric field that is excited by a laser in the plasma is when $\omega_{pe}=\omega_0$. When $\omega_{pe}>\omega_0$ the laser frequency is below the plasma electron characteristic frequency and driver is detuned. For the relativistic electron laser regime accessible with the present laser technology to be able to trap and accelerate ions, such high plasma fields are required. Hence, the accelerating structures for ions are excited at the critical density and ions are accelerated starting from rest. The condition on the excited ponderomotive potential to accelerate ions is $eZ_{ion}\phi_{pm}> (\gamma^{trap}_{ion}-1)M_{ion}c^2$. For this condition to be satisfied and for the laser-plasma ion accelerators to produce beams of kinetic energy comparable to laser-plasma electron accelerators the magnitude of $\phi^{ion}_{pm}$ has to be at least 3 orders of magnitude higher than the $\phi^{e}_{pm}$ as $M_{ion}/Z_{ion}>10^3m_e$. This is the reason that ion accelerators operate at highest possible plasma density because $-\vec{\nabla}\phi_{pm}=\vec{E}_{pm}\propto\sqrt{n_e}$ (for the same laser wavelength). It may be possible to create high enough electron plasma wave fields at high-$\beta$ to trap and accelerate relativistic ions but this needs hypothetical high-intensity lasers or lasers with x-ray wavelengths to access higher plasma densities and thereby higher plasma fields. However higher densities lead to smaller spatial scales of the accelerating structures in the plasma and the usable acceleration bucket sizes are smaller resulting in lower beam-charge. Such lasers can also directly lead to ion motion, which results in modification of the electron-ion displacement fields. 


\section{Critical layer in relation to other propagating plasma structures}
\label{section:plasma acc-structure}

In plasma-based electron acceleration mechanisms, the acceleration structure is a propagating plasma electron density wave at close to the speed of light, high-$\beta$. The plasma electron density wave is excited by electromagnetic energy or particle beams in their wake. The energy coupled into the plasma is used to excite traveling density structures with longitudinal and transverse space-charge fields in the frame of the propagating energy. These fields are used to transport and accelerate beams. In plasma electron accelerators the potential of the acceleration structure is limited by the ponderomotively or beam-field displaced electron density relative to the stationary background plasma-ions\cite{laser-driven}.

Because the electrons are lighter than the ions by three orders of magnitude, the potential required to relativistically transport and accelerate them is smaller. As discussed above, ions cannot be accelerated with relativistically propagating plasma structures, unless these density structures can create potentials high enough to trap ions. The maximum fields that are excited in a plasma are of the order of $\sqrt{n_e}$, because electron-ion space charge separation depends upon the charge density and the distance of separation inversely on density. So, higher densities create higher fields. Thus low-$\beta$ acceleration structures in the plasma are required to utilize the limited space-charge potential in the plasma that are excited by the accessible relativistic-electron momentum Ti:Sapphire, 800nm laser technology. The laser frequency limits the plasma density $n_e$ at which the laser fields can excite collective motion of plasma. If arbitrarily high frequency laser is available then the laser excited fields would be limited by the highest density material or density compression processes. In the low-$\beta$ limit, $\beta^{trap} \simeq \frac{1}{c}\sqrt{\frac{2e\phi_{pm}}{M_{ion}/Z_{ion}}}$ for trapping and accelerating ions.

The bulk plasma motion due to laser irradiation at the critical layer has been studied and utilized even in the non-relativistic laser pulse regime to create propagating collisional shock fronts. The collisional shock is a local density inflation (compression) structure excited by collisional (over the relevant timescales of different heating mechanisms) momentum transfer to plasma-ions from the electrons driven by photon momentum conservation of reflecting laser at the critical layer. Shock is sustained because the density inflation excited at the driven end is faster than the sound density-wave speed in the medium. The plasma density perturbations (not just plasma electrons) in the plasma cannot vibrationally (mean particle position is negligibly perturbed) propagate faster than the speed of sound. The supersonic density inflation on the driven side is not communicated across the shock. This results in the piling up of plasma density on the driven side against the cold plasma ahead of the shock. The formation of a shock is not dependent upon the laser imparting relativistic momentum to the plasma electrons. The shock front velocity is controlled to compress plasma to a high-density such as in direct-driven laser compression of fuel to facilitate fusion and ignition of the whole fuel. The laser power is spread over a large focal spot size because the fuel plasma has to be uniformly compressed over the whole fuel volume to achieve consistently high density in the fuel bulk (to avoid in-homogeneities in compression) with as few laser beams as possible. So, the laser intensities in this scenario never get close to relativistic amplitudes and are of the order $\simeq 10^{14}\frac{W}{cm^2}$. At these intensities the shock speeds are of the order $10^{-3}$c. Hence these shocks cannot accelerate relativistic particle beams. In this regime the lasers pulse lengths tend to be of the order of nanoseconds to launch multiple shocks to compress the material for long enough duration to achieve a high confinement time. At such intensities and pulse-lengths the laser energy is of the order of tens of kilo-Joules. So, such bulky expensive lasers and the propagating structures at these intensities cannot be readily viable as a source of relativistic particle beams for wide ranging applications.

However, there is technologically a more accessible regime of near-infra-red ultra-short laser pulses of 10s of femto-seconds which have energy around 100 Joules. Such lasers are based upon the Ti:Sapphire ($\lambda_0=800nm$) active medium technology and can be setup on a table-top with the current state-of-the-art (size and cost are pump laser size and technology limited). When the light from these accessible lasers is focussed to a spot-size of the order of a wavelength the intensities reach in excess of $10^{22}W-cm^{-2}$. The processes that are addressed with these lasers are in the regime where the interacting electrons gain relativistic momentum directly in the laser field. At such high intensities any target is readily ionized into a plasma (metals are ablated into plasma for intensities $>10^{15}W-cm^{-2}$) and thereby plasma electrons control the dynamics of the interaction. At the wavelengths of Ti:Sapphire laser the critical density, $n_e^{crit}=m_e(2\pi c/\lambda_0)^2/(4\pi e^2)$ is $\sim 2\times 10^{21}cm^{-3}$. We further analyze the interaction of these very high intensity ultrashort pulse lasers with targets that have densities greater than critical density. These lasers have a significant rise-time (large fraction of their duration) from amplified spontaneous emission intensity pedestal to peak intensity. There is also a fraction of laser energy in the pre-pulse which precedes the main pulse. The target ablation starts in the pedestal or pre-pulse and the relativistic intensity portion of the laser interacts with the pre-plasma. 

By controlling the motion of the critical layer through relativistically induced transparency we can excite low-$\beta$ propagating plasma accelerating structures. The relativistic electron momentum gained by the plasma electrons in the strong laser field leads to an apparent relativistic increase in their mass by the Lorentz factor $\gamma_e(r,t) = \sqrt{1 + \vec{p}_e(r,t).\vec{p}_e(r,t) /m_e^2c^2}$. With this space-time varying relativistic correction the plasma frequency is reduced and the laser can propagate to the relativistic plasma critical density, $n^{\gamma_e}_{crit} = \gamma_e \frac{m_e^{\gamma_e(r,t)\simeq 1}}{4\pi e^2}$. In the first order approximate analysis we infer the propagation of a relativistic laser to its relativistic corrected critical density. However, the laser ponderomotive force on the plasma electrons creates a large density inflation (snowplow) ahead of the laser. The higher than the local electron density in the ponderomotively excited snowplow results in a modification of the plasma frequency and the laser pulse cannot propagate to its relativistically corrected critical density.

In addition to the plasma critical layer motion, such low-$\beta$ propagating plasma accelerating structures are excited by the free expansion of the plasma into vacuum \cite{schreiber-2006}. This effect is utilized in the Target Normal Sheath Acceleration (TNSA) scheme. In this scheme the fast electrons driven into a plasma vacuum interface are trapped by the electrostatic pull of the immobile ions at the interface. The trapped fast electrons then thermalize and exchange energy with the ions. The heated plasma sheath then expands to equilibrate the thermal pressure of the hot electrons and creates a propagating space-charge field \cite{mora-2003}. The thermal pressure of the plasma thereby determines the rate of the expansion of the sheath. The fast electrons are accelerated by a laser ponderomotively driving the plasma electrons in a pre-plasma towards the plasma-vacuum interface or from an external injector. 

Similarly, exciting the plasma critical layer using a laser pulse can create a collision-less shock front that can propagate at controlled speeds. The shock primarily differs from RPA in that it is freely propagating and is not continuously driven by the laser. A critical layer electron layer of high longitudinal momentum is excited by  reflected laser photon momentum conservation. The exchange of momentum to the plasma ions is through collision-less mechanism (electron mean free path is greater than the Debye length) mediated by the space-charge potential of the displaced electron layer. The primary energy dissipation for such a shock is by reflection of the plasma ahead of it. When the electron layer space-charge potential is above a threshold it electrostatically drags the plasma-ions to supersonic velocity. The shock front is not inflicted by transverse instabilities seeded by laser filamentation. Transverse and longitudinal stability of the shock front depends upon its Mach number. Thermal excitation of the plasma ions by the laser is important because this leads to a higher speed of sound in the plasma which leads to lower Mach numbers for a fixed bulk plasma shock velocity. A shock propagating with low Mach numbers is laminar, where different transverse fluid elements are coherent. High Mach number shocks tend to be turbulent which leads to poor reflected beam quality. 

\begin{figure*}
	\begin{center}
   	\includegraphics[width=6.8in]{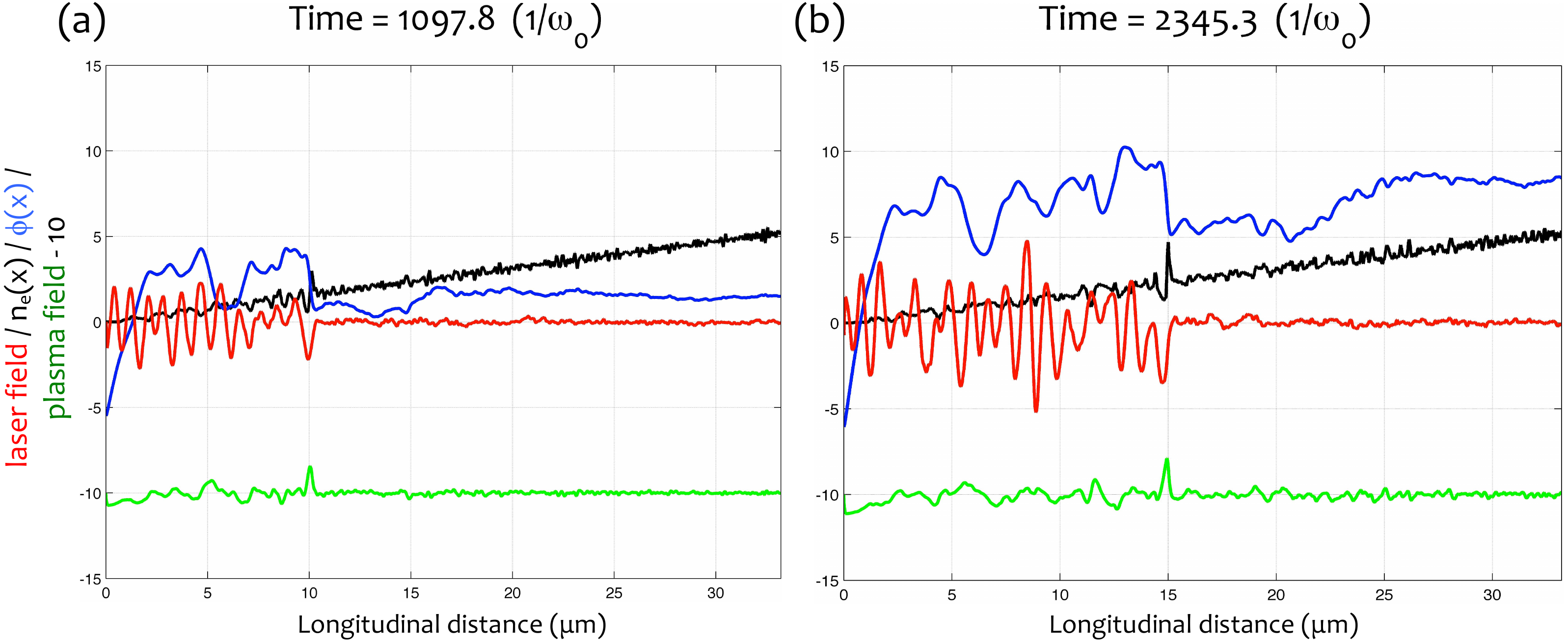}
	\end{center}
\caption{ { \scriptsize {\bf 1-D PIC simulations with fixed background plasma ions in continuously rising density with $\alpha=50\frac{c}{\omega_0}$ interacting with continuously rising laser of $a_0=2$, $\delta=1000\frac{c}{\omega_0}$}. This plot shows in normalized units, plasma electron density (black, normalized to the critical density), laser transverse electric field (red), plasma space-charge longitudinal field (green, scaled by $-10$) and ponderomotively driven potential (in blue). The RITA laser front moves forward into a fixed-ion plasma density gradient with densities higher than the critical density. Initial critical density is located at $x=50\frac{c}{\omega_0}=6.4\mu m$. At $t=1100\frac{1}{\omega_0}$ (in (a)) the relativistic critical layer is at $x=10\mu m$ (for $a(x,t)\simeq 2.2$) and at a later time $t=2345\frac{1}{\omega_0}$ (in (b)) it is at $x=15\mu m$ (for $a(x,t)\simeq 3.5$). The simulations are unlike Fig.\ref{fig:heavy-ion-plasma-concept} as the laser is continuously rising for the entire simulation (no flat or falling part to terminate the forward motion). } }
\label{fig:heavy-ion-plasma-1d-snapshot}
\end{figure*}

\begin{figure*}
	\begin{center}
   	\includegraphics[width=6.8in]{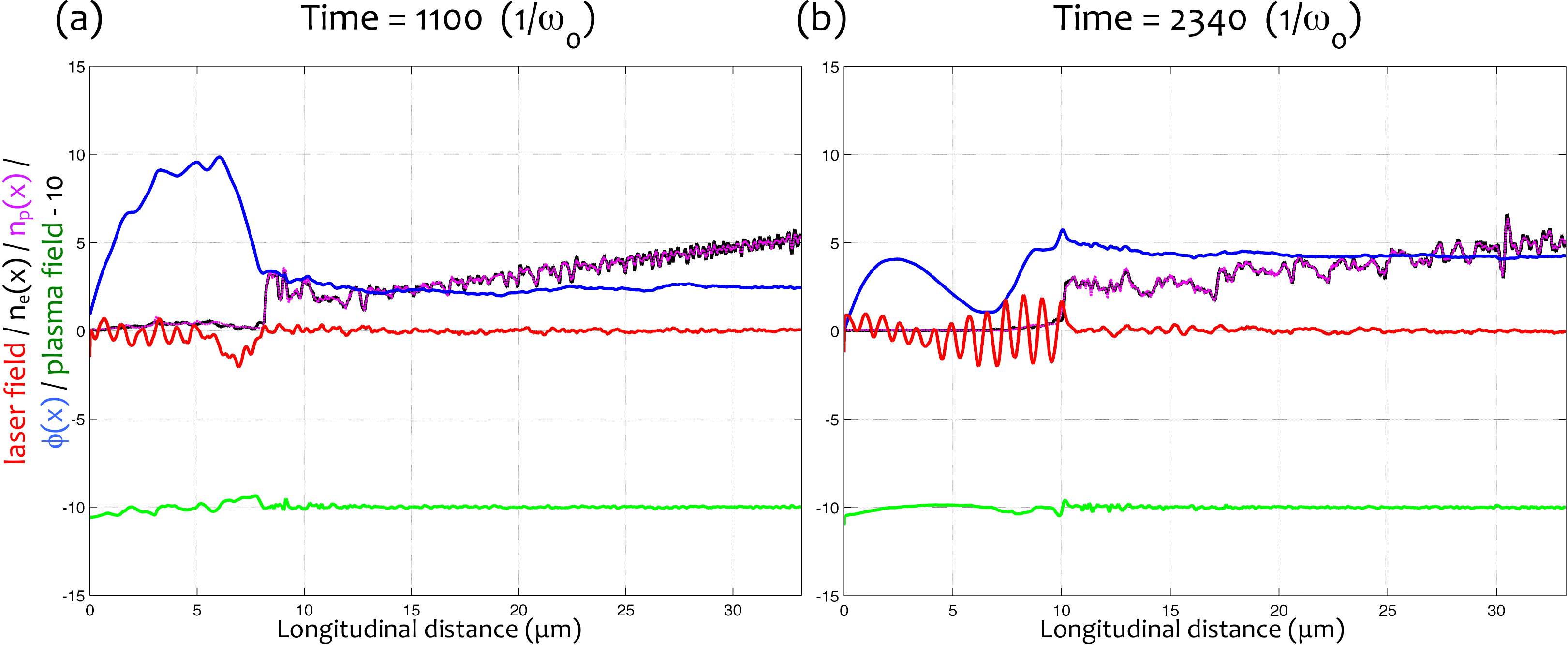}
	\end{center}
\caption{ { \scriptsize {\bf 1-D PIC simulations in a light co-moving plasma ($M_{ion}=m_p=1836m_e$) of longitudinally rising plasma density with $\alpha=50\frac{c}{\omega_0}$ and an infinitely long flat-top laser pulse intensity profile with $a_0=2$}. The color scheme is as in Fig.\ref{fig:heavy-ion-plasma-1d-snapshot} with plasma-ion density in magenta. Even though the laser field amplitude is constant it is able to propagate beyond the initial critical density. At $t=1100\frac{1}{\omega_0}$ (in (a)) the plasma critical layer observed at the laser front has moved to $x=8\mu m$ and at a later time at $t=2340\frac{1}{\omega_0}$ (in (b)) along with the whole plasma it has moved to $x=10\mu m$. On comparing with heavy-ion plasma in Fig.\ref{fig:heavy-ion-plasma-1d-snapshot}, the plasma electron density is significantly modified only in the regions around the relativistic critical layer. Whereas with light plasma ions the plasma density in the region just behind the critical layer is evacuated, resembling a hole.} }
\label{fig:light-ion-plasma-1d-snapshot}
\end{figure*}



\section{Analysis of the plasma critical layer motion}
\label{sec:crit-layer motion analysis}

In the following subsections \ref{subsec:heavy-ion model} and \ref{subsec:light-ion model}, we analytically demonstrate the key difference in the physical processes of relativistically induced transparency and radiation pressure. By a review of the analytical models developed in \cite{RITA-PRE-2013} and \cite{hole-boring-1992}, we show that different mechanisms occur in heavy-ion and light co-moving ion critical layer laser-plasma interactions leading to different scaling laws.

In the subsection \ref{subsec:neglected-effects-in-model} we present the physical effects that have been ignored in the 1-D analytical model. The sections on 2-D results therefore cannot be accurately compared to the 1-D analytical model. 

\subsection{Heavy fixed plasma ions - critical layer motion 1-D model}
\label{subsec:heavy-ion model}

The motion of the plasma critical layer when the plasma-ions are stationary is analyzed. This model uses results developed in \cite{Kaw-Dawson}\cite{Akhiezer-Polovin}. We model 1-D interaction of ultra-short relativistic-electron laser with heavy-ion plasma assuming that the plasma ions are stationary. To analyze the critical layer dynamics during the relativistic laser-plasma interactions we only consider plasma electron motion at the critical layer. The plasma density is inhomogeneous, $n_e^{`}(x)$. To the first order of approximation we assume that the plasma density is invariant during the interaction. The laser pulse envelope $a^2(x,t)$ is changing with time (assumed not changing during the interaction) and propagating forward into the plasma by inducing transparency relativistically. In free-space the laser envelope is characterized in the frame $f(x-ct)$. During the rising edge of the laser pulse an increment in the relativistic plasma electron momentum leads to the motion of the laser front to the relativistically corrected critical layer. A rising plasma density provides control over the velocity of the laser packet. The laser front velocity is controlled in a rising plasma density if the relativistically corrected critical layer is ahead of the location of the laser at any instant.

The increasing plasma frequency in the spatially rising plasma density gradient in 1-D is $\omega^2_{pe}(x)=4\pi n^{`}_e(x)e^2/m_e$. When the laser amplitude is low such that the momentum $\vec{p}_e$ is negligible compared to the $m_ec$, critical density is defined as $n^{\gamma_e\simeq 1}_{crit}(x_{crit}) = \omega_0^2 m_e / 4\pi e^2$. The rising intensity of the head of the laser pulse excites increasing relativistic momentum of the plasma electrons at the location of the laser field interacting with the critical layer electrons. The rising laser envelope field thereby increases the momentum of the quivering plasma electrons such that it becomes significant and then surpasses the $m_ec$. Since, the plasma frequency is modified by the laser field exciting relativistic momentum $\gamma_e(x,t)\vec{\beta}_e=\vec{a}(x,t)$ to the electrons, the plasma frequency of the critical layer irradiated by the rising laser intensity becomes $(\omega^{\gamma_e}_{pe}(x,t))^2=1/\gamma_e(x,t)\left(4\pi n^{`}_e(x)e^2/m_e\right)$. The relativistic critical layer at each instant of interaction, $t$, is defined by the relativistically corrected plasma frequency equalling the laser pulse frequency $\omega_{pe}^{\gamma_e}(x,t)=\omega_0$. The relativistic critical layer therefore has a characteristic plasma electron collective frequency that can reflect the laser. When the plasma electrons are not relativistic the laser reflects at $n_{crit}^{\gamma_e\simeq 1}$. The relativistic critical layer is defined at the plasma electron density where the effective density has collective oscillations to shield the laser field $n_e(x)/\gamma_e(x,t)=n_{crit}^{\gamma_e\simeq 1}$. The term layer refers to the laser reflection density contour in full-space. At each instant the laser is reflected off from the relativistic critical layer which itself moves in the laser propagation direction. To determine the rate of the forward motion of the critical layer due to relativistic effect, we partially differentiate $(\omega^{\gamma_e}_{pe}(x,t))^2 = \omega_0^2\frac{n_e(x)}{\gamma_e(x,t)}$ with $t$ and $x$ (where, $n_e(x)=n^{`}_e(x)/n^{\gamma_e\simeq 1}_{crit}=\omega^2_{pe}(x)/\omega_0^2$). When partially differentiating with $t$ we have the time variation, $\frac{\partial (\omega^{\gamma_e}_{pe}(x,t))^2}{\partial t} = \omega_0^2 n_e(x) \frac{\partial}{\partial t}\gamma^{-1}_e(x,t)$. When partially differentiating with $x$ we have the 1-D spatial variation as, $\frac{\partial (\omega^{\gamma_e}_{pe}(x,t))^2}{\partial x} = \omega_0^2 \left( \gamma^{-1}_e(x,t) \frac{\partial}{\partial x} n_e(x) + n_e(x) \frac{\partial}{\partial x}\gamma^{-1}_e(x,t) \right) $. We assume that the laser pulse frequency is fixed and therefore $\frac{\partial \omega_0(x,t)}{\partial x} =  \frac{\partial \omega_0(x,t)}{\partial t} = 0$\cite{ChITA-2012}. The product of the time derivative with the inverse of the spatial derivative gives the speed of the relativistic critical layer, $\frac{\partial (\omega^{\gamma_e}_{pe}(x,t))^2}{\partial t} \left(\frac{\partial (\omega^{\gamma_e}_{pe}(x,t))^2}{\partial x} \right)^{-1} =  \frac{\partial x}{\partial t} = v_{sp}(x,t) = n_e(x) \frac{\partial}{\partial t}\gamma^{-1}_e(x,t) \left( \gamma^{-1}_e(x,t) \frac{\partial}{\partial x} n_e(x) + n_e(x) \frac{\partial}{\partial x}\gamma^{-1}_e(x,t) \right)^{-1} $. Since the quiver motion is first order in $\vec{a}(x,t)$ and directly excites all plasma electrons, we can ignore the second order effects resulting from the ponderomotive force ($\propto \nabla |\vec{a}|^2(x,t)$). Under the conservation of the canonical momentum, $\vec{p}_e^{\perp}=\vec{a}(x,t)$. From PIC simulations we observe that the ponderomotive force (in the density gradient) does not excite the plasma electrons coherently in comparison with the first order transverse quiver oscillations. Therefore we can assume that the leading order contribution to the apparent relativistic mass increase by the Lorentz factor is due to the direct Lorentz force excited quiver motion of the plasma electrons in the laser field. So, the Lorentz factor is $\gamma_e(x,t)\simeq\gamma_e^{\perp}(x,t)=\sqrt{1+\vec{p}_e^{\perp}(x,t).\vec{p}_e^{\perp}(x,t)/m_e^2c^2}=\sqrt{1+|\vec{a}(x,t)|^2}$. With this Lorentz factor, $\frac{\partial}{\partial t} \gamma^{-1}_e(x,t) = -\frac{1}{2} \gamma^{-3}_e(x,t) \frac{\partial}{\partial t}a^2(x,t)$ and $\frac{\partial }{\partial x} \gamma^{-1}_e(x,t) = -\frac{1}{2} \gamma^{-3}_e(x,t) \frac{\partial }{\partial x} a^2(x,t)$. Using these partial derivatives of the Lorentz factor, $v_{sp}(x,t) =  \frac{1}{2} n_e(x) \frac{\partial}{\partial t}a^2(x,t) \left(  \gamma^2_e(x,t) \frac{\partial}{\partial x} n_e(x)  -  \frac{1}{2} n_e(x)\frac{\partial}{\partial x}a^2(x,t)  \right)^{-1} $.

To simplify the analysis we assume linear models for $a^2(x,t)$ and $n_e(x)$. In order to characterize the linear models we choose the scale-lengths as the parameters. The uphill plasma density rises linearly with scale-length $\alpha$, as $n_e(x) = n_{cold}^{crit}\left(\frac{x}{\alpha}\right)$. When $x=\alpha$ the rising plasma density reaches the cold-plasma non-relativistic critical density. For each step of the longitudinal distance along $x$ by $m\alpha$ the plasma density rises to $mn_{cold}^{crit}$. The laser pulse intensity also rises linearly, with the rise-time scale length $\delta$ (rise-time = $\frac{\delta}{c}$), as $a^2(x,t) = a_0^2\left(\frac{ct-x}{\delta}\right)H(ct-x)$ (where H is the Heaviside step function). There is a first order approximation here that the laser pulse shape is not significantly changed in the plasma. At any location $x$ where the head of the laser pulse has already reached, the pulse envelope rises linearly in time (pulse envelope is referenced to the $ct-x$ which is in the causal region of the rest-frame of the laser).  For each time-step of $m\delta/c$ the laser intensity rises to $ma_0^2$. Using these linear models, we have, $\frac{\partial}{\partial t}a^2(x,t) = a_0^2\frac{c}{\delta}$ and $\frac{\partial}{\partial x}a^2(x,t) = -a_0^2\frac{1}{\delta}$. The first-order approximation of stationarity of plasma density assumes a negligible perturbation is driven by the second order laser force. With these expressions we can calculate the velocity of the relativistic critical layer as, $v_{sp} = -\frac{1}{2}n_e(x)a_0^2\frac{c}{\delta}\left( -\frac{1}{2}n_e(x)a_0^2\frac{1}{\delta} - (\gamma_e(x,t))^2 n_{crit} /\alpha \right)^{-1}$. Since the relativistic plasma critical layer always satisfies the condition that $n_e(x)=\gamma_e(x,t)n_{crit}$. We can re-write the velocity as $v_{sp}(t)/c =  \frac{1}{2}a_0^2\frac{\alpha}{\delta}  \left(\frac{n_e(x)}{\gamma_e(x,t)n_{crit}}\right)  \left( \frac{1}{2} a_0^2\frac{\alpha}{\delta} \left(\frac{n_e(x)}{\gamma_e(x,t)n_{crit}}\right)  + \gamma_e(x,t) \right)^{-1} =  \frac{1}{2}a_0^2\frac{\alpha}{\delta}  \left( \frac{1}{2} a_0^2\frac{\alpha}{\delta}  + \gamma_e(x,t) \right)^{-1}$. 

The above 1-D model does not take into account a few realistic physical processes. These are briefly explained in \ref{subsec:neglected-effects-in-model} and arise due to higher-order effects of laser-plasma interactions. The RITA model is valid only under the condition that $v_{sp}(t)/c\le v_{g}(t)/c<1$. To understand the model we can try to simplify the expression under the assumption that $ \frac{1}{2} a_0^2\frac{\alpha}{\delta}  \ll \gamma_e(x,t)$. With this assumption the snowplow velocity expression simplifies to $v_{sp}(t)/c = \frac{1}{2} a_0\frac{\alpha}{\delta}$ and this scaling law model has been verified for small $a_0$ in \cite{RITA-PRE-2013}. It is shown using simulations that the simplified model presented above quantifies the physical process of relativistically induced transparency. 

\begin{figure}
	\begin{center}
   	\includegraphics[width=3.4in]{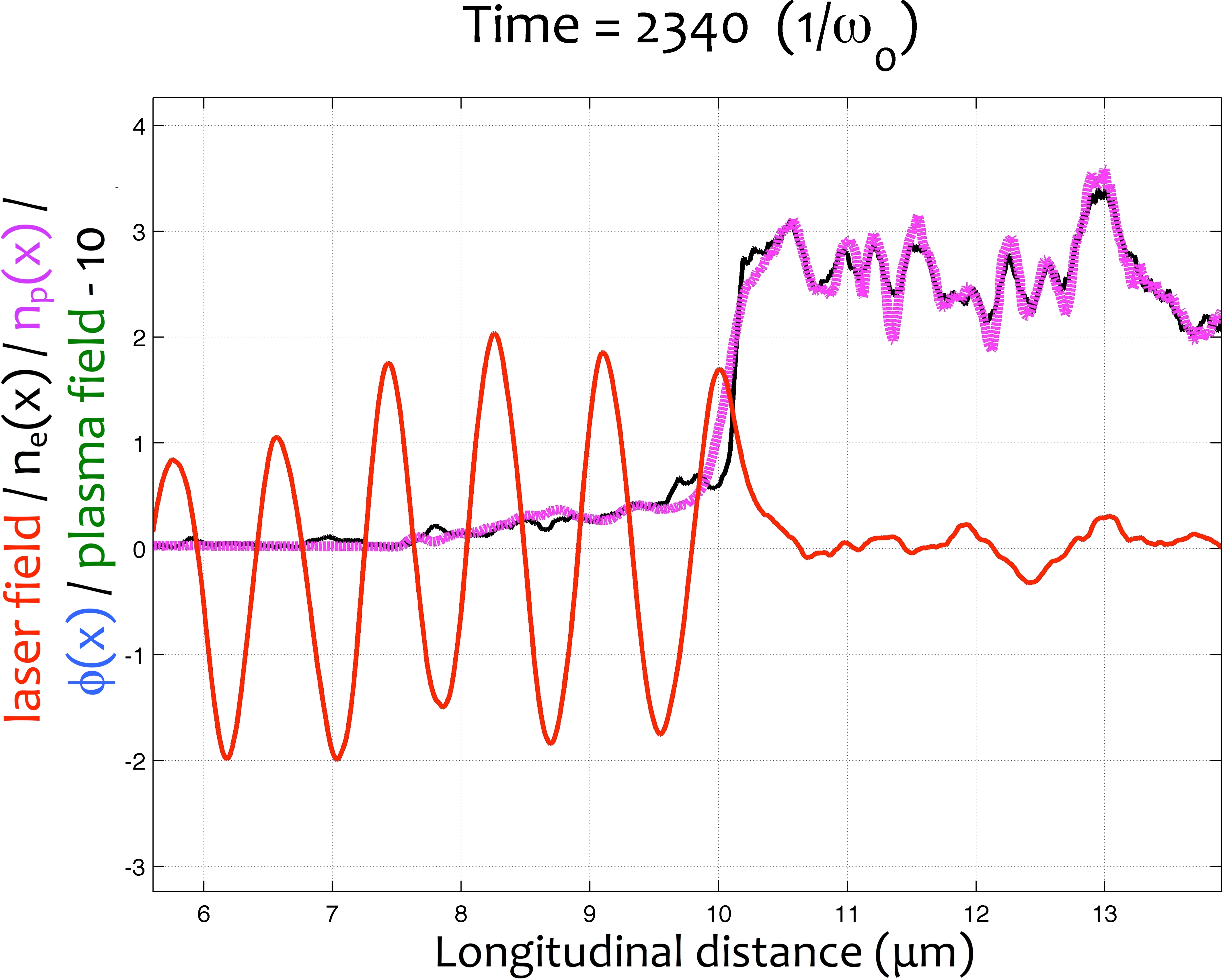}
	\end{center}
\caption{ { \scriptsize {\bf Zoomed in critical layer in 1-D PIC line plot of radiation pressure driven plasma ($M_{ion}=m_p=1836m_e$)}. In a zoomed-in view of the critical layer in Fig.\ref{fig:light-ion-plasma-1d-snapshot}(b), the plasma ions (magenta) co-move with the plasma electron layer (black). The plasma electron layer is longitudinally ahead of the ion layer. The reflection point, beyond which the laser field (red) drops exponentially in amplitude is approximately at $n_e(x) \simeq 1.5$. The laser being relativistic reflects at the relativistic critical density.} }
\label{fig:close-up-hole-boring}
\end{figure}

\begin{figure*}
	\begin{center}
   	\includegraphics[width=6.8in]{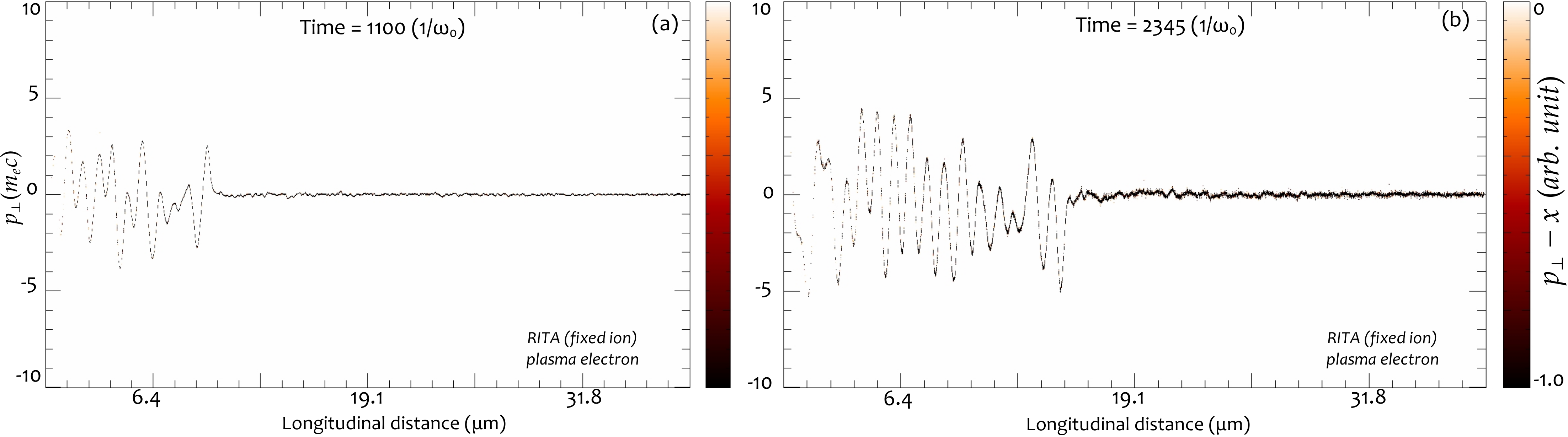}
	\end{center}
\caption{ { \scriptsize {\bf Phase-space (1-D PIC) of transverse electron ($p_{\perp}$) momentum-density product in fixed ion plasma with laser-plasma parameters in Fig.\ref{fig:heavy-ion-plasma-1d-snapshot}}. The transverse momentum ($p_{\perp}(x,t)=a(x,t)$) is coherently excited for all electrons collocated with the laser up to the longitudinal position of the laser front. The location of the relativistic critical layer corresponds to Fig.\ref{fig:heavy-ion-plasma-1d-snapshot}(a) and (b) respectively. At the critical layer in (a) at $t=1100\frac{1}{\omega_0}$, $p_{\perp} \simeq 2$ and in (b) at $t=2345\frac{1}{\omega_0}$, $p_{\perp} \simeq 3.5$. Therefore as the laser field $a(x,t)$ rises with time, $p_{\perp}$ at the laser front critical layer increases. Coherent increase in the transverse momentum of all irradiated plasma electrons leads to relativistically induced transparency of the laser.} }
\label{fig:p2x1-heavy-ion-plasma-1d-snapshot}
\end{figure*}

\begin{figure*}
	\begin{center}
   	\includegraphics[width=6.8in]{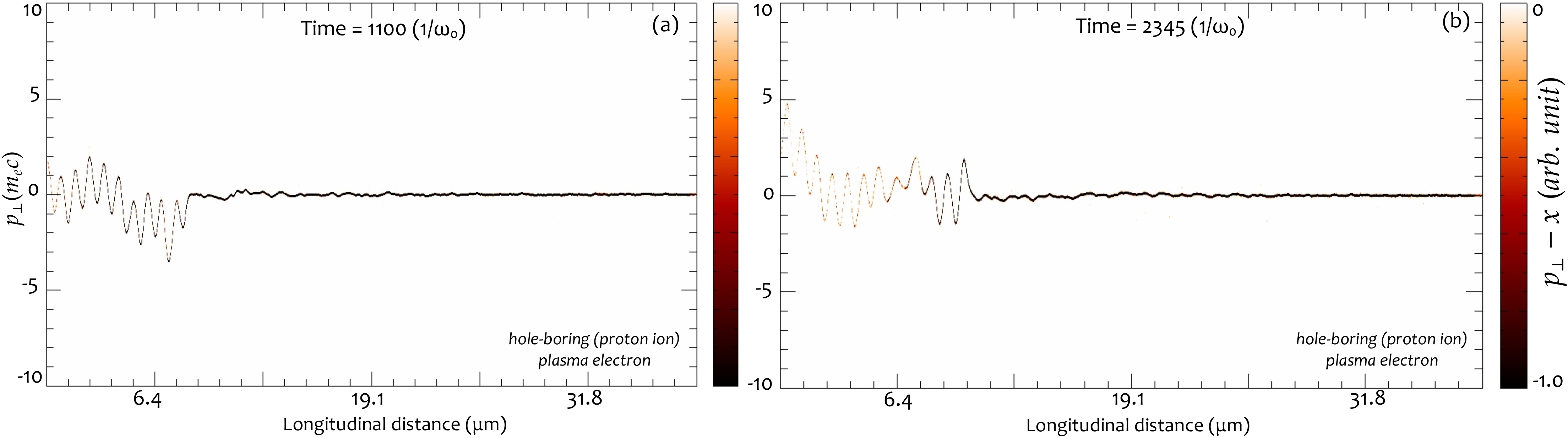}
	\end{center}
\caption{ { \scriptsize {\bf Phase-space (1-D PIC) of transverse electron ($p_{\perp}$) momentum-density product in a light co-moving ion plasma with laser-plasma parameters in Fig.\ref{fig:light-ion-plasma-1d-snapshot}}. From (a) and (b) it is observed that $p_{\perp}$ corresponding to Fig.\ref{fig:light-ion-plasma-1d-snapshot}(a) and (b) is $a_0 = 2$. The maximum $p_{\perp}$ remains a constant due to a constant laser intensity. The modulation in $p_{\perp}$ is due to the interference between the normally incident laser pulse and the Doppler-shifted reflected radiation from the critical layer. } }
\label{fig:p2x1-light-ion-plasma-1d-snapshot}
\end{figure*}

\begin{figure*}
	\begin{center}
   	\includegraphics[width=6.8in]{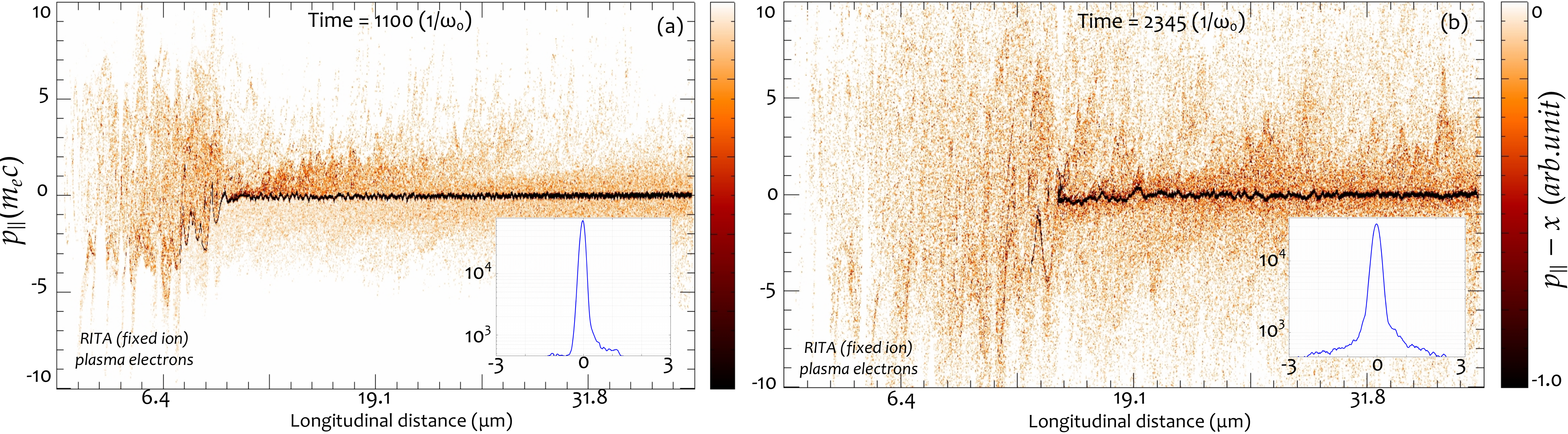}
	\end{center}
\caption{ { \scriptsize {\bf Phase space (1-D PIC) of electron longitudinal ($p_{\parallel}$) momentum-density product in a fixed background-ion plasma with laser-plasma parameters in Fig.\ref{fig:heavy-ion-plasma-1d-snapshot}}. The longitudinal momentum of the plasma electrons at relativistic critical layer as in Fig.\ref{fig:heavy-ion-plasma-1d-snapshot}. In comparison to $p_{\perp}$ Fig.\ref{fig:p2x1-heavy-ion-plasma-1d-snapshot} $p_{\parallel}$ is not coherent. The forward momentum population is observed in the electron momentum spectra (inset). There is significant forward electron flux that escape the pull of the background ions. A return current is setup as the ponderomotive electron reach the boundary as in (b).} }
\label{fig:p1x1-heavy-ion-plasma-1d-snapshot}
\end{figure*}

\begin{figure*}
	\begin{center}
   	\includegraphics[width=6.8in]{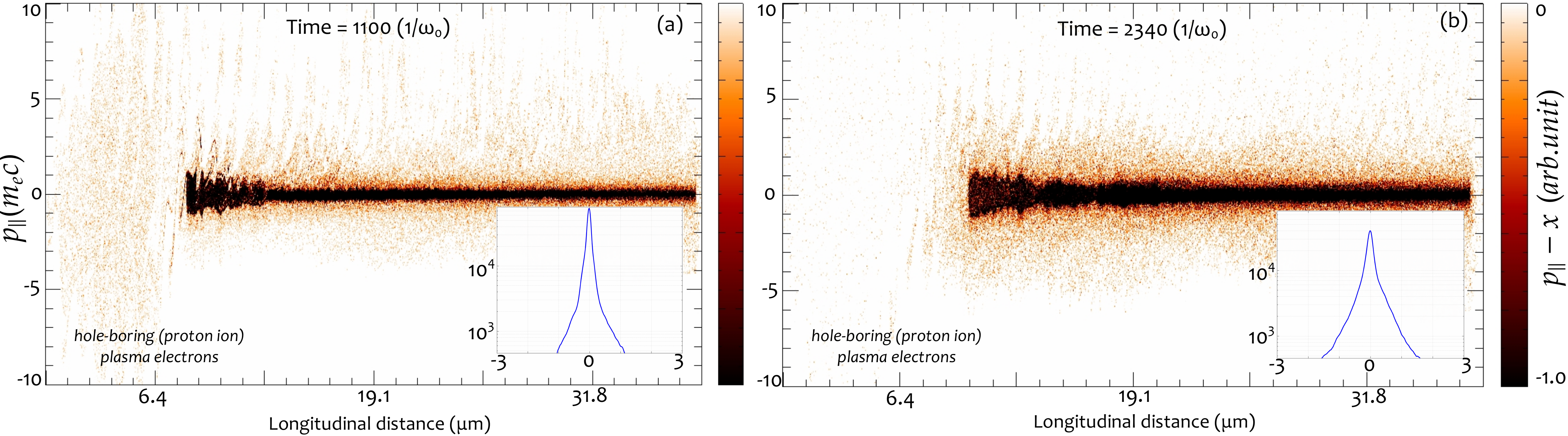}
	\end{center}
\caption{ { \scriptsize {\bf Phase space (1-D PIC) longitudinal ($p_{\parallel}$) electron momentum-density product in a light co-moving ion plasma with laser-plasma parameters in Fig.\ref{fig:light-ion-plasma-1d-snapshot}}. The longitudinal momentum of the plasma electrons at critical layer as in Fig.\ref{fig:light-ion-plasma-1d-snapshot}. It is sen this case the electron longitudinal momentum spectra (inset) is symmetric. This is because the plasma electrons transfer significant longitudinal momentum to the background plasma-ions resulting in the co-propagation of the plasma-ions. As a result the electrons longitudinally oscillate around the co-moving plasma-ions instead of escaping the background ion electrostatic field and gaining only a forward momentum.} }
\label{fig:p1x1-light-ion-plasma-1d-snapshot}
\end{figure*}

\subsection{Co-moving plasma ions - critical layer motion 1-D model}
\label{subsec:light-ion model}

We now analyze the motion of the plasma critical layer when the plasma-ions move with the laser front. The plasma critical layer acts as a reflecting mirror to the incident radiation. The conservation of momentum ($3^{rd}$ law of motion) of the reflected radiation results in the momentum transfer to the plasma, the critical layer is driven forward in the direction of the laser incidence. The velocity of the forward motion of the radiation pressure driven plasma is estimated by equating the energy density of the radiation to the momentum flux density rate of the whole plasma. 

This model follows the analysis presented in \cite{hole-boring-1992} (HB) and \cite{rpa-2004} (LS). The model proposes that all the plasma ions at density $n_{pi}$ are forced to move with the radiation pressure front. It is evident that there is an interplay of two forces, one originating from the photon momentum reversal and the second due to the space-charge force of the displaced electron layer. Hence, there is a threshold when the bulk plasma motion is possible. In the frame of the moving laser front \cite{hole-boring-1992} (HB) by equating the momentum flux of mass flow $\left(\frac{M_{ion}}{Z_{ion}}v_{HB}(t)\right) n_{pi} v_{HB}(t)$ to the light pressure $2I(t)/c$ (for complete reflection $|\rho|\simeq 1$) we get, $v_{HB}(t)/c=\sqrt{ \frac{2I(t)}{n_{pi}c^3M_{ion}/Z_{ion}} }$. Using, $n_{crit} = \frac{\omega_0^2m_e}{4\pi e^2}$ and $I(t)=\frac{\pi c}{2} \left( \frac{m_e c^2}{e\lambda_0} \right)^2 a(t)^2$, the speed of the motion of the laser front is $v_{HB}(t)/c=\left( \frac{ n_{crit} }{ n_{e}(x) } \frac{ Z_{ion} m_e }{ M_{ion} } a^2(t) \right)^{1/2}$. When the critical layer front reflecting the laser is propagating at $v_{HB}(t)$, the laser reflects off a relativistically moving surface. Under certain relativistic corrections the velocity of the critical layer slows down\cite{rel-HB-2009}. 

Realistic laser pulses do not have a flat-top intensity profile and their intensity envelope significantly varies with time. Similarly, the plasma density is not homogeneous due to the formation of the pre-plasma. Inhomogeneous plasma density is encountered by the rising relativistic intensity laser pulse due to the pre-ionization of a homogeneous density target during the laser pre-pulse and the diffusive expansion of the pre-plasma. Hence, the velocity of the hole-boring radiation pressure acceleration front $v_{HB}(t)$ is not constant during the laser pulse envelope interacting with inhomogeneous plasma density. There is a threshold intensity of the laser pulse above which the radiation pressure is high enough to electrostatically pull the plasma ions. When the laser intensity above the threshold changes significantly, the velocity of the hole-boring front also changes correspondingly. However, if the laser pulse envelope is modified such as to control the rise in intensity while the plasma density increases, a certain control over the hole-boring velocity may be achieved.

\subsection{Physical effects neglected - plasma critical layer motion 1-D model}
\label{subsec:neglected-effects-in-model}

These simplified models presented above do not account for many physical effects, some prominent ones are detailed below.

{\it Self-focussing in reducing skin-depth} - In a longitudinally increasing plasma density the plasma skin depth decreases. As a result the laser experiences a self-focussing \cite{self-focussing} force of the plasma which is increasing. A reducing radius channel is formed due to the balance between the transverse laser ponderomotive force and the plasma-ion electrostatic force holding back the ponderomotive electrons. However in relativistic laser-plasma interactions the plasma frequency of relativistic plasma electrons is suppressed by the square root of the Lorentz factor of relativistic longitudinal density compression (or relativistic mass increase). During interaction with relativistic-electron laser the plasma skin-depth is modified to $\sqrt{\gamma_e}\frac{c}{\omega_p}$ as a result the self-focussing characteristics are different from non-relativistically excited plasma. So, transverse instability is suppressed by $\sqrt{\gamma_e}$. The value of $a^2(x,t)$ and $I(t)$ dynamically increase as the laser transverse envelope experiences self-focussing from the ponderomotive plasma electron channel. The increase in the laser intensity due to self-focussing is a 2-D effect as it affects the transverse profile of the laser and is not accounted for in the model of the laser front velocity in \ref{subsec:heavy-ion model} and \ref{subsec:light-ion model}. However, it should be noted that the intensity rise-time and thereby the acceleration length are extended during the self-focussing effect.

{\it Incident intensity modulation due to interference with reflected light} - When the laser pulse is incident on a reflecting surface the reflected laser pulse propagating in the opposite direction to incidence interferes with the incident pulse. The interference from a stationary reflecting surface does not result in a beat pattern of high modulation depth. However, when the critical layer is relativistically moving, the reflected laser pulse undergoes a Doppler shift. The Doppler shifted reflected light results in a beat pattern on interfering with the incident pulse envelope. The effective incident laser pulse $a(x,t)$ is therefore modulated and the laser pulse amplitude is not as incident. The effect of the Doppler-shifted reflected radiation is highest for small angles of incidence to the normal. This modulation of the laser pulse envelope by the Doppler-shifted reflection is not accounted for in the model of the laser front velocity in \ref{subsec:heavy-ion model} and \ref{subsec:light-ion model}. This is similar to the process of ponderomotive bunching as explored in \cite{ponderomotive-force}.

{\it Laser intensity and frequency transformation in relativistic frame of the critical layer} - When the moving critical layer (located within the acceleration structure) reaches relativistic velocity there is a significant modification of the incident laser frequency in the frame of the propagating critical layer. This is due to relative velocity between the laser and the acceleration structure in the frame of reference of the critical layer. This results in red-shift of the laser in the frame co-moving with the critical layer at the laser front. The redshift of the laser pulse in the frame of the moving critical layer is characterized by the z-parameter as in astronomy, $z+1 = \frac{\lambda_{obs}}{\lambda_0} = \frac{f_0}{f_{obs}}$. The observed frequency of the laser pulse in the moving critical layer frame is $\omega_{obs} = \omega_0 / (z+1) = \omega_0 \sqrt{\frac{1-\beta}{1+\beta}}$. Because the laser pulse undergoes a red-shift, the critical layer recedes behind $n_{crit}^{red-shift}=\frac{\omega_{obs}^2m_e}{4\pi e^2} = \frac{\omega_0^2m_e}{4\pi e^2} \left(\frac{1-\beta}{1+\beta}\right)$. As a result the critical layer forward propagation slows down because its velocity depends upon the laser intensity (which is frequency dependent) in both the models in \ref{subsec:heavy-ion model} and \ref{subsec:light-ion model}. The ratio $I_0/\omega_0^3$ is a relativistic invariant \cite{intensity-doppler-1985}.
 
{\it Ponderomotive electron build-up at the critical layer} - For laser pulses at very high intensities with fixed pulse duration, the rate of the rise of intensity is high. As a result the ponderomotive force due to a higher rate of rise of the laser intensity on the plasma electrons at the critical layer is higher. This results in accretion of the plasma electrons just ahead of the critical layer. This non-equilibrium non-quasi-neutral electron density inflation is significantly higher compared to the neutralizing background heavy-ion density. In heavy-ion plasma this electron density inflation leads to modification of the plasma electron frequency within the electron inflation. As the electron density inflation results in the increase in the plasma frequency just ahead of the laser front, it suppresses the induced transparency. The exact modification of the plasma frequency due to the propagating local electron density inflation which is not in equilibrium with background plasma ions is beyond the scope of this work. However, electron build up has been analyzed in a homogeneous plasma with fixed ions in a stationary critical layer \cite{striction}.

{\it Ion motion and beam-loading in relativistically induced transparency propagation} - As the acceleration structure transfers momentum to the accelerated beam of ions it loses energy and the snowplow speed and potential drop. This is an effect that can slow down the electron snowplow speed. However, the effect of beam-loading is not considered in \ref{subsec:heavy-ion model}. In \cite{RITA-PRE-2013} an increase in the trace proton density leads to a lower peak energy but a narrower peak in the energy spectra. Additionally, while the acceleration structure potential is high enough to accelerate the trace density protons it is likely to still have an impact on the background plasma heavy-ions. However, the speed of the plasma acceleration structure is high enough that the momentum transfer to the heavy-ions is limited by the interaction time, $\Delta p^{\parallel}_{heavy-ion} = F^{\Delta}_{sp}\times\Delta t^{\parallel} = e \nabla\phi_{sp}\times\frac{\Delta x^{\parallel}}{v_{sp}}$. In the model presented in \ref{subsec:heavy-ion model} this effect is not accounted for. If the speed of the acceleration structure is scaled down then the energy loss to the background heavy-ions is scaled up as $\Delta p_{heavy-ion} \propto \frac{\nabla\phi_{sp}}{v_{sp}}$.

\begin{figure*}
	\begin{center}
   	\includegraphics[width=6.8in]{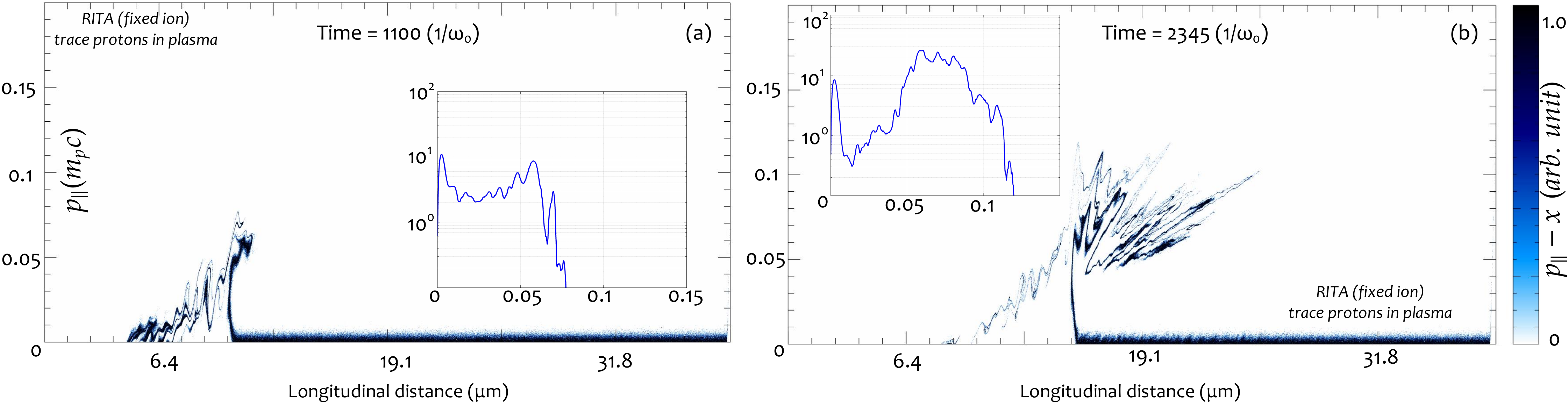}
	\end{center}
\caption{ { \scriptsize {\bf Phase-space (1-D PIC) of accelerated trace-density proton longitudinal ($p_{\parallel}$) momentum-density product with fixed background ion plasma with laser-plasma parameters in Fig.\ref{fig:heavy-ion-plasma-1d-snapshot}}. The trace-density protons ($m_p=1836m_e$) are introduced a low density of $0.01n_{crit}$ and do not affect the dynamics. The longitudinal momenta correspond to the instant and location same as in Fig.\ref{fig:heavy-ion-plasma-1d-snapshot}. In (a) trace protons are launched at twice the speed of the relativistic critical layer (0.041c from \ref{subsec:heavy-ion model} and average speed of 0.032c from Fig.\ref{fig:heavy-ion-plasma-1d-snapshot}). The reflected proton bunch has a momentum around $0.064c$. A large fraction of the trace protons are reflected by the snowplow potential as observed from the proton momentum spectrum (inset). } }
\label{fig:proton-p1x1-heavy-ion-plasma-1d-sim}
\end{figure*}

\begin{figure*}
	\begin{center}
   	\includegraphics[width=6.8in]{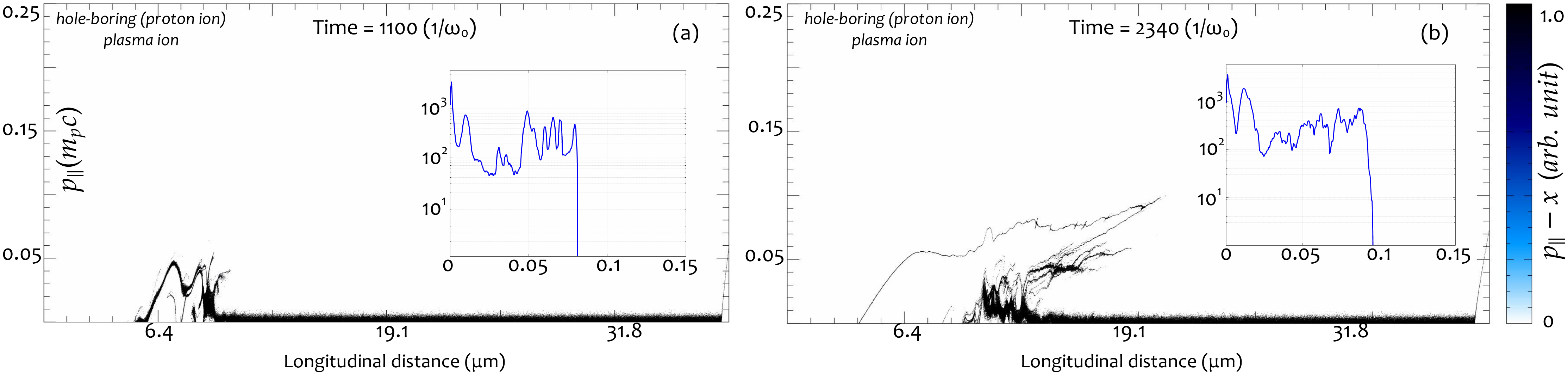}
	\end{center}
\caption{ { \scriptsize {\bf Phase space (1-D PIC) of longitudinal ($p_{\parallel}$) plasma-ion (proton) momentum-density product in light co-moving ion plasma with laser-plasma parameters in Fig.\ref{fig:light-ion-plasma-1d-snapshot}}. These phase-spaces correspond to plasma-ion momentum in Fig.\ref{fig:light-ion-plasma-1d-snapshot}. A high fraction of the protons in (a) are at the same momentum as the speed of the hole-boring laser front (0.047c from \ref{subsec:light-ion model} with average speed 0.023c from Fig.\ref{fig:light-ion-plasma-1d-snapshot}). However, in (b) we observe a small fraction of the protons also being launched at twice the velocity of the hole-boring laser front. In comparison to the proton momentum spectra in Fig.\ref{fig:proton-p1x1-heavy-ion-plasma-1d-sim} the accelerated proton density is 2 orders of magnitude higher in hole-boring because in RITA the accelerated species is a doped trace species at $\simeq 0.01n_{crit}$. The proton momentum spectra (inset) has a large spread because of the plasma density gradient. } }
\label{fig:proton-p1x1-light-ion-plasma-1d-sim}
\end{figure*}


\section{1-D simulation of the plasma critical layer motion}
\label{sec:crit-layer-motion-1d-sim}

Using 1-D PIC simulations we seek to answer the primary question whether forward motion of the relativistic plasma critical layer can occur while the plasma ions are infinitely massive. This investigates the process of relativistically induced transparency. Also, using 1-D PIC simulations we analyze the interaction of a flat-top laser pulse profile with a light co-moving ion (proton) plasma which answers whether the plasma-ions co-propagates with the laser front due to the radiation pressure of conservation of photon momentum reversal at the critical layer.

The PIC simulations\cite{osiris-code-2002} use normalized units, where distance is normalized to $\frac{c}{\omega_{0}}$ and time is normalized to $\frac{1}{\omega_{0}}$. The density is normalized to the critical density so it is unity where plasma frequency equals the laser frequency, $\omega_0=\omega_{pe}$. We use $40\pi$ grid cells per laser wavelength and 40 particles per species per cell. The electrons and protons each have an initial temperature of 5keV. For heavy-ion plasma we initialize with an immobile neutralizing background. In light co-moving ion plasma simulation we initialize with plasma ions as proton, $m_p=1836m_e$. The simulation domain is 400$\frac{c}{\omega_0}$ long and the linear plasma density gradient scale length is $\alpha=50\frac{c}{\omega_0}$. To convert from simulation units to real units we take $\lambda_0=0.8\mu m$ for comparison to Ti:Sapphire laser based experiments. However, for collision-less plasmas all simulations can be scaled keeping the same value of $\frac{\omega_{pe}}{\omega_0}$. We use open boundary conditions for fields and particles and third-order particle shapes with current smoothing and compensation. 

We find in 1-D simulations with fixed background plasma ions to model heavy-ion plasma that the motion of the relativistic critical layer is by the process of relativistically induced transparency. We setup the laser to have a linearly rising envelope with a rise time scale-length of $\delta=1000\frac{c}{\omega_0}$ and the normalized laser vector potential of $a_0=2$. So every $\Delta t=1000\frac{1}{\omega_0}$ the laser intensity ($a^2(x,t)$, squared field amplitude) at the left boundary (where the laser is launched) increases by an integer. In another simulation, not shown here, with a flat-top laser pulse there is no forward motion. 

The propagation of the relativistic critical layer to different longitudinal positions is in Fig.\ref{fig:heavy-ion-plasma-1d-snapshot}(a) and (b). The plasma density is not evacuated behind the relativistic critical layer as the fixed background ions drag back the electrons when the laser amplitude undergoes a temporal minima during the beat interference pattern. The forward motion of the relativistic critical layer is attributed to the rising Lorentz factor in the rising laser field amplitude. We observe rising coherent quiver transverse momentum ($p_{\perp}$ normalized to $m_ec$) of all the plasma electrons from the under dense (left boundary to $50\frac{c}{\omega_0}$) region up to the location of the relativistic critical layer in Fig.\ref{fig:p2x1-heavy-ion-plasma-1d-snapshot}(a) and (b). It is observed that the relativistic critical layer moves forward due to the rising coherent relativistic quiver momentum of the plasma electrons. Therefore rising relativistic field amplitude of the laser is a necessary condition for the forward propagation of the critical layer in a plasma with fixed-ions. The ponderomotively excited longitudinal momentum $p_{\parallel}$ of the quivering electron (only a small fraction of electrons) is not factored in to the Lorentz factor for the suppression of the plasma frequency because as we can see from Fig.\ref{fig:p1x1-heavy-ion-plasma-1d-snapshot} longitudinal oscillations are not coherent. Due to the ponderomotive force being a second-order force, plasma $p_{\parallel}$ in the density gradient is stochastic. As a result a fraction of ponderomotive electrons escape the critical layer space-charge while others are collocated or execute longitudinal oscillations. The distribution is biased towards forward momenta as in the electron momenta spectrum in Fig.\ref{fig:p1x1-heavy-ion-plasma-1d-snapshot} inset. At a later time in Fig.\ref{fig:p1x1-heavy-ion-plasma-1d-snapshot}(b) we can also observe a return current in the $p_{\parallel}$ phase-space as in $p_{\parallel}<0$ inset spectra.

To demonstrate that the snowplow of electrons and its space-charge ponderomotive potential excited at the relativistic critical layer is used to accelerate ions, we initialize a trace proton species at $0.01n_{crit}$ density. The longitudinal momentum phase-space of these trace density protons is in Fig.\ref{fig:proton-p1x1-heavy-ion-plasma-1d-sim}. From the simulation snapshots in Fig.\ref{fig:heavy-ion-plasma-1d-snapshot} we observe that the speed of the relativistic critical layer is $\simeq0.032c$ in comparison to $v_{sp}=\frac{1}{2}2^2\frac{50}{1000} / (\frac{1}{2}2^2\frac{50}{1000} + \sqrt{1 + 2^2})=0.41c$. The accelerated trace-density proton momentum is distributed around $0.064m_pc$ (in Fig.\ref{fig:proton-p1x1-heavy-ion-plasma-1d-sim}(a)). At the later time Fig.\ref{fig:proton-p1x1-heavy-ion-plasma-1d-sim}(b), we observe a large fraction of the trace-protons at $\simeq0.8m_pc$ (closer to $2 \times 0.41c$). The acceleration of the trace-density protons occurs due to the reflection off the snowplow potential of the trace-density protons ahead of the relativistic critical layer as explained in \cite{RITA-PRE-2013}.

In plasma with light co-moving ions (protons) interacting with a flat-top laser pulse we observe from 1-D simulations that whole plasma moves forward longitudinally and the critical layer moves forward with it. This is evident in Fig.\ref{fig:light-ion-plasma-1d-snapshot}(a) and (b). In these simulations the laser intensity profile is constant with a normalized vector potential of $a_0=2$. The laser has a rise-time of $10\frac{1}{\omega_0}$ (little over one cycle). The radiation pressure of the long flat-top laser reflecting off the critical layer keeps pushing the whole plasma density forward. The electron layer is pushed forward by the laser radiation pressure and its electrostatic potential drags along the plasma ions, a zoomed in view of this interaction at the critical layer is in Fig.\ref{fig:close-up-hole-boring}. Relativistic quiver is not a necessary condition for the forward propagation of the radiation pressure driven critical layer in light co-moving ion plasma. 

In Fig.\ref{fig:p2x1-light-ion-plasma-1d-snapshot} the constant amplitude of the flat-top laser quivers the plasma electrons at the critical layer to $p_{\perp}=a_0$. Because the light-ions co-propagate longitudinal forward with the plasma electrons, the longitudinal momentum in this case is symmetric (in Fig.\ref{fig:p1x1-light-ion-plasma-1d-snapshot} and inset $p_{\parallel}$ momentum spectra). From this we can infer that the plasma electrons undergo longitudinal oscillations ($\omega_{pe}$ or \cite{fast-osc-2004}) in the frame of the laser front in addition to an average forward propagation. The plasma-ion longitudinal momentum phase-space is in Fig.\ref{fig:proton-p1x1-light-ion-plasma-1d-sim}. The light plasma-ions co-propagate with the electrons at the laser front critical layer and therefore pick up a longitudinal momentum $\simeq0.045m_pc$ equal to the velocity of the hole-boring front. With $v_{HB}=\sqrt{\frac{2(8.56 \times 10^{18})}{c(2.1*10^{21})(1.67262178 \times 10^{-27})}} = \sqrt{\frac{4}{1836.15}}c = 0.047c$ from \ref{subsec:light-ion model}.  However, in Fig.\ref{fig:proton-p1x1-light-ion-plasma-1d-sim}(b) at a later time we also observe a small fraction of protons with longitudinal momentum at around twice the hole-boring velocity.

Another critical difference is in the longitudinal electric field (green curved scaled by -10) noticed by comparing Fig.\ref{fig:heavy-ion-plasma-1d-snapshot} and Fig.\ref{fig:light-ion-plasma-1d-snapshot}. The magnitude and shape of the field are different due to the difference in the background ion spatial distribution behind the electron layers.

Therefore, from simulations it is observed that the plasma critical layer can propagate forward while the plasma ions are infinitely massive. This longitudinal motion of the critical layer requires increasing the laser field amplitude $a_0>1$ to propagate by relativistically induced transparency. Also, the plasma critical layer can propagate along with the whole plasma when the plasma ions co-move with the electron layer. The radiation pressure based forward motion does not necessarily require relativistic amplitude laser. 


\begin{figure*}
	\begin{center}
   	\includegraphics[width=6.8in]{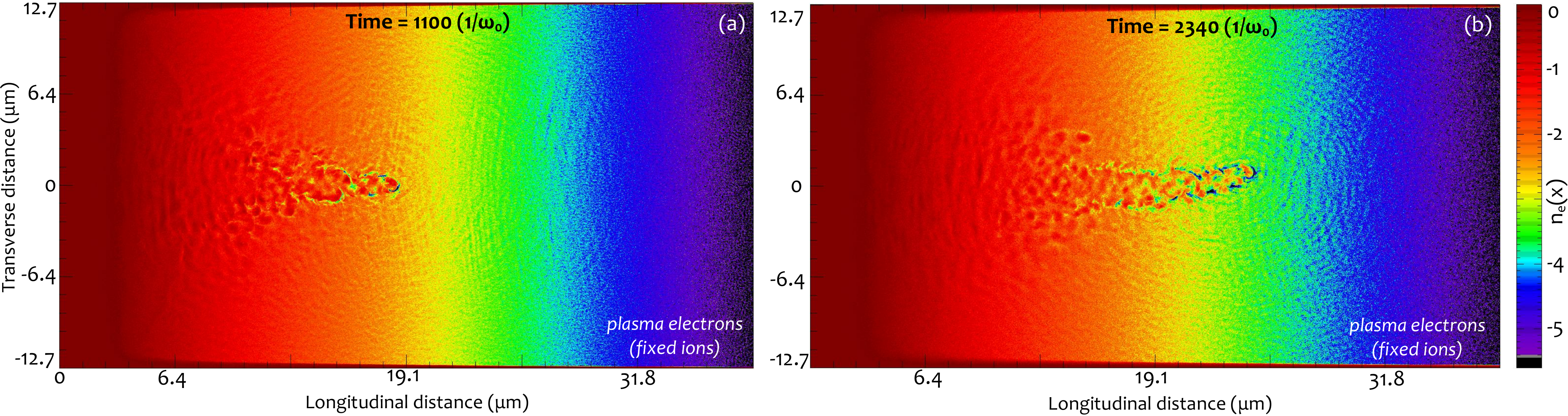}
	\end{center}
\caption{ { \scriptsize {\bf 2-D PIC electron density in real-space in fixed heavy background ion plasma density gradient of $\alpha=50\frac{c}{\omega_0}=6.4\mu m$ interacting with continuously rising laser of $a_0=2$, $\delta=1000\frac{c}{\omega_0}$ and a laser FWHM with radius $r_0=3.8\mu m$. The density rises to $\simeq 5.6 n_{crit}$ between $2.6\mu m$ and $38.2\mu m$}. (a) and (b) are at the same instants as in 1-D Fig.\ref{fig:heavy-ion-plasma-1d-snapshot}. In (a) at $t=1100\frac{1}{\omega_0}$ the relativistic critical layer is at $\simeq 16\mu m$ (density gradient starts at $2.6\mu m$) compared to $10\mu m$ in the 1-D Fig.\ref{fig:heavy-ion-plasma-1d-snapshot}(a) and in (b) at $\simeq 22\mu m$ as compared to $15\mu m$ in Fig.\ref{fig:heavy-ion-plasma-1d-snapshot}(b). This is due to the self-focussing of the laser pulse increasing $a(x,t)$ in the reducing $\frac{c}{\omega_p(x)}$. Region behind the relativistic critical layer is not evacuated. This is because the fixed-ions in the background equilibrate the ponderomotively excited electrons.} }
\label{fig:2D-heavy-ion-plasma-electrons}
\end{figure*}

\begin{figure*}
	\begin{center}
   	\includegraphics[width=6.8in]{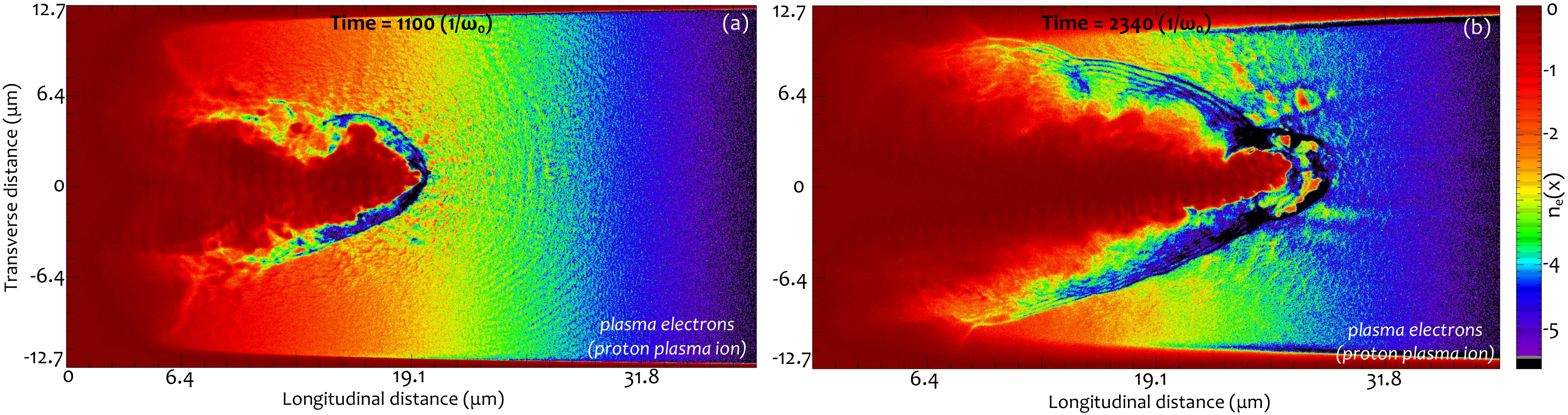}
	\end{center}
\caption{ { \scriptsize {\bf 2-D PIC electron density in real-space in co-moving light (proton $m_p=1836m_e$) background ion plasma density gradient of $\alpha=50\frac{c}{\omega_0}=6.4\mu m$ interacting with an infinitely long flat laser of $a_0=2$ and focal FWHM $r_0=3.8\mu m$.}. In comparison with 2-D Fig.\ref{fig:2D-heavy-ion-plasma-electrons} there is a significant difference in the propagation of the critical layer due to protons as the background ions. In (a) the hole-boring laser front is at $18\mu m$ and in (b) it is at $24\mu m$. In 2-D the hole-boring front is considerably faster in comparison with 1-D simulations in Fig.\ref{fig:light-ion-plasma-1d-snapshot}. } }
\label{fig:2D-light-ion-plasma-electrons}
\end{figure*}

\begin{figure*}
	\begin{center}
   	\includegraphics[width=6.8in]{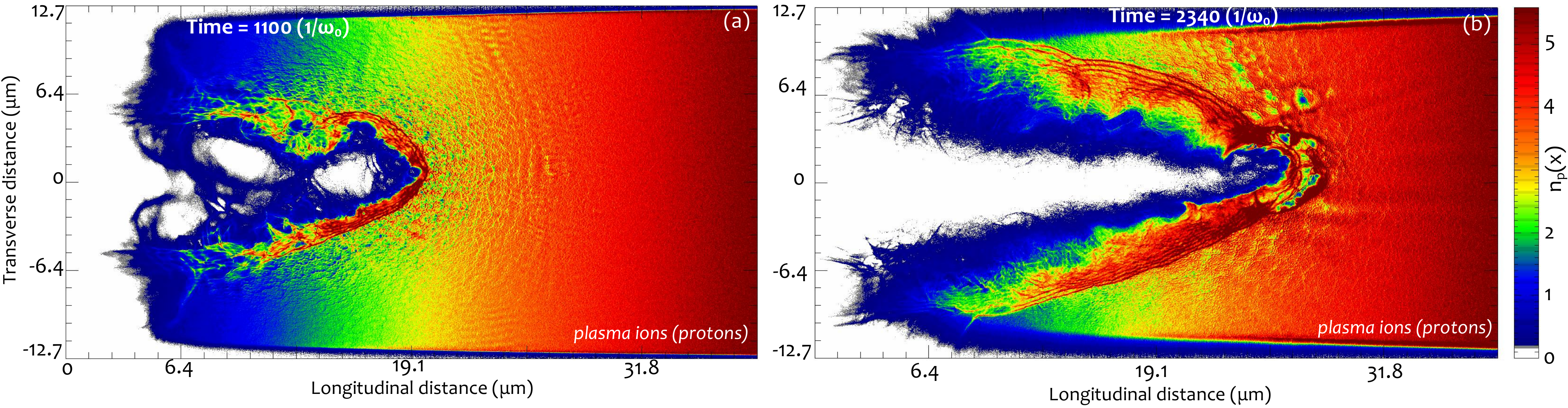}
	\end{center}
\caption{ { \scriptsize {\bf 2-D PIC background-ion in real-space in a plasma with co-moving light background ions ($m_p=1836m_e$) for the same laser-plasma parameters as in \ref{fig:2D-light-ion-plasma-electrons}.} The plasma-ion density is at the same instants as in Fig.\ref{fig:2D-light-ion-plasma-electrons}. The plasma ion density co-propagates with the electrons density and develops density perturbations similar in amplitude and spatial profile as the plasma electron density. In contrast to the Fig.\ref{fig:2D-heavy-ion-plasma-electrons} the self-focussed plasma channel has larger transverse extent. Because the plasma ions are co-moving with the electrons, a weaker space-charge electrostatic pull is exerted to balance the transverse ponderomotive force on the electrons.} }
\label{fig:2D-light-ion-plasma-protons}
\end{figure*}

\begin{figure*}
	\begin{center}
   	\includegraphics[width=6.8in]{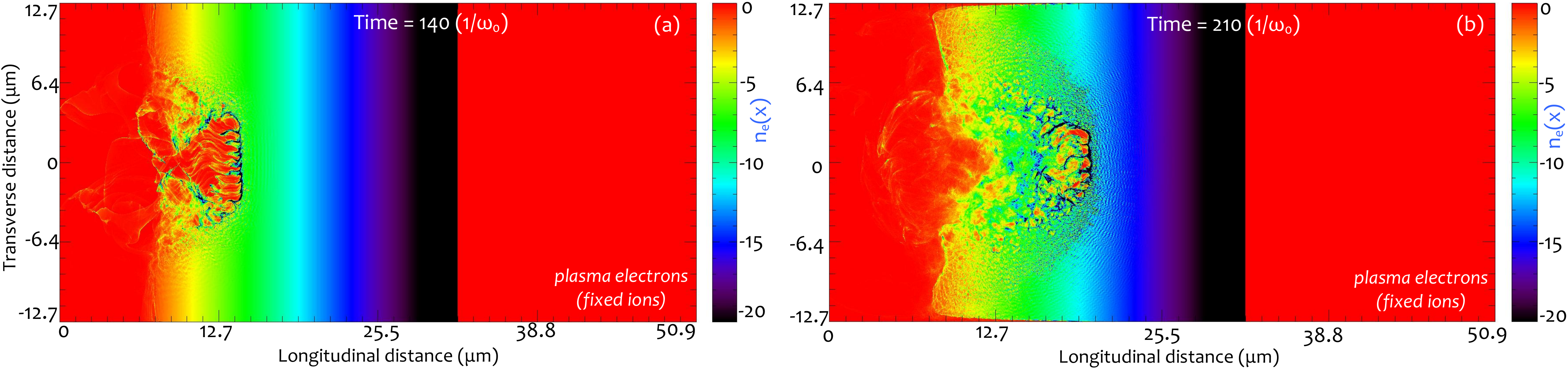}
	\end{center}
\caption{ { \scriptsize {\bf 2-D PIC electron density in real-space in a heavy-ion plasma density gradient of $\alpha=8.8\frac{c}{\omega_0}=1.1\mu m$ interacting with laser $I_0=1.1\times 10^{22}\frac{W}{cm^2}$ ($a_0=72$, $r_0=3.8\mu m$)}. In (a) at $t=140\frac{1}{\omega_0}$ the relativistic critical layer is at $14\mu m$ and in (b) at $t=210\frac{1}{\omega_0}$ it stops around $20.4\mu m$. The relativistic critical layer stops much ahead of the expected plasma density $n_e^{\gamma}\simeq72$ up to where a relativistic laser of $a_0=72$ would propagate in theory. The suppression of relativistic critical density is due to the second order effect of the plasma electron build-up (\ref{subsec:neglected-effects-in-model}) or electron snowplow steepening at the head of the laser due to the ponderomotive force.} }
\label{fig:2D-heavy-ion-electrons-high-intensity}
\end{figure*}


\section{2-D simulations of the plasma critical layer motion}
\label{sec:crit-layer-motion-2d-sim} 

The 2-D PIC simulations use the same normalization and $20\pi$ grid cells are used per laser wavelength $\lambda_0$. We use 36 particles per species, with 6 particles in each of the dimension. The simulation domain is generally 400$\frac{c}{\omega_0}$ long and 200$\frac{c}{\omega_0}$ wide. The transverse spatial profile of the laser normalized vector potential, $\vec{a}$ is chosen to be Gaussian, $a(\vec{r})=a~exp(\frac{-r^2}{r_0^2})\hat{r}$ (intensity is super-Gaussian). In all the 2-D simulations laser focal FWHM radius is $r_0=30\frac{c}{\omega_0}=3.8\mu m$. The parameter $a_0$ is converted to laser power using $I=|\vec{a}|^2\left(\frac{\pi c}{2}\right)\left(\frac{m_ec}{e\lambda}\right)^2$, ($I_{peak}=I_0$), its focal spot-size radius $r_0$ and its peak power $P_0=\pi r_0^2\frac{I_0}{2}$. The electrons, trace protons and the background-ions have an initial thermal distribution with average temperature of 5keV. The particles and fields in the simulation have open boundaries and particles have third-order shapes with current smoothing and compensation. The frequency of the laser pulse in these simulations is not chirped \cite{ChITA-2012}.

The laser-plasma parameters to study the motion of the plasma critical layer in heavy and light co-moving ion plasma are exactly the same as described in 1-D simulations (section \ref{sec:crit-layer-motion-1d-sim}). The motion of plasma critical layer similar to 1-D simulations is observed. In 2-D simulations of fixed-ion plasma forward motion of the relativistic critical layer again occurs by induced transparency as in the electron density snapshots in Fig.\ref{fig:2D-heavy-ion-plasma-electrons}. In 2-D simulations of the light co-moving ion plasma in Fig.\ref{fig:2D-light-ion-plasma-electrons} we see co-propagation of plasma-ions in Fig.\ref{fig:2D-light-ion-plasma-protons} with the electrons. In the region behind the laser front from where the background light-ions are evacuated, up to where the electron layer peaks we do not observer a sharp boundary (has a finite density ramp) unlike in the 1-D simulations. This finite gradient ramp behind the layer is in Fig.\ref{fig:2D-light-ion-plasma-electrons}(b). Also, there is a significant difference from the 1-D simulations in the self-focussing effect on the laser pulse transverse envelope and the increase in $a(x,t)$ due to the reducing transverse spot size as it propagates into the plasma. The self-focussing is evident because the laser propagates in a rising plasma density gradient and the skin-depth reduces, $\frac{c}{\omega_p(x)}$ (\ref{subsec:neglected-effects-in-model}). Therefore the plasma density gradient acts as a channel to control the increase in the laser intensity by relativistic ponderomotive self-focussing.

There is also a significant difference in the self-focussing of the laser pulse in heavy-ion and the light ion-plasma, observed by comparing Fig.\ref{fig:2D-heavy-ion-plasma-electrons} and Fig.\ref{fig:2D-light-ion-plasma-electrons}. In the plasma with light co-moving ions the laser self-focuses into a larger transverse spatial size channel in Fig.\ref{fig:2D-light-ion-plasma-electrons}. In comparison the laser focusses into a much narrow channel in heavy-ion plasma. This physical effect arises because in the light co-moving ion plasma the plasma-ions which provide the space-charge electrostatic force to compensate the ponderomotive force on the plasma electrons co-propagate. Thereby the magnitude of space-charge force on the ponderomotively excited electrons is lowered in comparison with stationary ions. In the fixed-ion plasma (or heavy-ion plasma) the ponderomotively excited plasma electrons that create the self-focussing channel experience a much larger space-charge force from the stationary plasma-ions. Therefore the channel in fixed-ion plasma in Fig.\ref{fig:2D-heavy-ion-plasma-electrons} is smaller in transverse size as result of much high charge separation force.


\section{Ultra-relativistic 2-D simulations of the plasma critical layer motion}
\label{sec:high intensity 2d-sim}

Motion of the plasma critical layer irradiated by ultra-relativistic intensity laser is compared in a heavy-ion to light co-moving ion (proton) plasma for the same laser-plasma parameters to analyze the two processes discussed above. In this section proton acceleration application of the plasma critical layer motion is emphasized for LIA by a GeV proton beam trapped and accelerated by the motion of the plasma critical layer.

In a plasma with fixed heavy background ions when the incident laser pulse has a peak intensity of $I_0=1.1\times 10^{22}\frac{W}{cm^2}$ for normalized laser vector potential of $a_0=72$ and a laser FWHM with radius $r_0=3.8\mu m$. The plasma density rises with $\alpha=8.8\frac{c}{\omega_0}=1.1\mu m$ and rises to $\simeq 22 n_e(x)$ in $25.5\mu m$. The plasma critical layer motion is in electron density dynamics in Fig.\ref{fig:2D-heavy-ion-electrons-high-intensity}. The trace protons at $0.01n_{crit}$ doped in the plasma are accelerated to around a GeV as in the phase space in Fig.\ref{fig:2D-heavy-ion-proton-high-intensity} (energy spectra inset). To study the effect of ion motion in RITA, realistic heavy plasma-ion is initialized instead of a uniform neutralizing background. When we use background plasma-ions of $M^*_{ion}=M_{ion}/Z_{ion}=10m_p$, the moving relativistic plasma critical layer transfers negligible momentum to the heavy-ions. The speed of the acceleration structure is not affected and neither is the potential scaled down. The maximum momentum transfer to the plasma heavy-ions occurs when the acceleration structure slows down before stopping as the interaction time of the potential with the plasma-ions increases. The density perturbations and longitudinal momentum phase space at $t=210\frac{1}{\omega_0}$ just before the snowplow stops are in Fig.\ref{fig:2D-heavy-ion-ions-high-intensity}(a) and (b) respectively. The simulation in Fig.\ref{fig:2D-heavy-ion-ions-high-intensity} is distinct from the fixed ion and is performed additionally to show the minimal impact of heavy-ions (instead of a neutralizing background).

At high laser intensities around $10^{22}\frac{W}{cm^2}$ the hole-boring radiation pressure acceleration beam energy estimation for flat-top sharply rising optimized plasma densities is around 300 MeV \cite{rel-HB-2009}\cite{HB-acc-2012}. We simulate the hole-boring plasma critical layer motion in a light co-moving ion (proton, $m_p=1836m_e$) plasma with the density gradient as described above. The plasma electrons and the plasma light-ions have the same order of density perturbations and co-propagate at the laser front, as in Fig.\ref{fig:2D-light-ion-high-intensity}(a) and (b). This clearly shows the formation of a double layer with co-propagating plasma-ions. However, the hole in the plasma electrons and ions is not completely evacuated. We also observe that the plasma protons achieve longitudinal momentum as high as the speed of the plasma critical layer as in Fig.\ref{fig:2D-light-ion-high-intensity}(c). 

The critical layer motion in a heavy-ion plasma with $M_{ion}=10m_p$ by the radiation pressure mechanism would result in the speed of the critical layer scaled down by the square root of the plasma-ion mass, $\frac{1}{\sqrt{10}} v^{proton}_{HB}$. Additionally, the ubiquitous pre-plasma density gradient modifies the $v^{proton}_{HB}$. In a comparison of the accelerated proton longitudinal momentum phase space in Fig.\ref{fig:2D-heavy-ion-proton-high-intensity} by radiation pressure mechanism to Fig.\ref{fig:2D-light-ion-high-intensity}(c) by relativistically induced transparency, the difference of the gain in the accelerated ion energy is not incremental.

\begin{figure}[h!]
	\begin{center}
   	\includegraphics[width=3.4in]{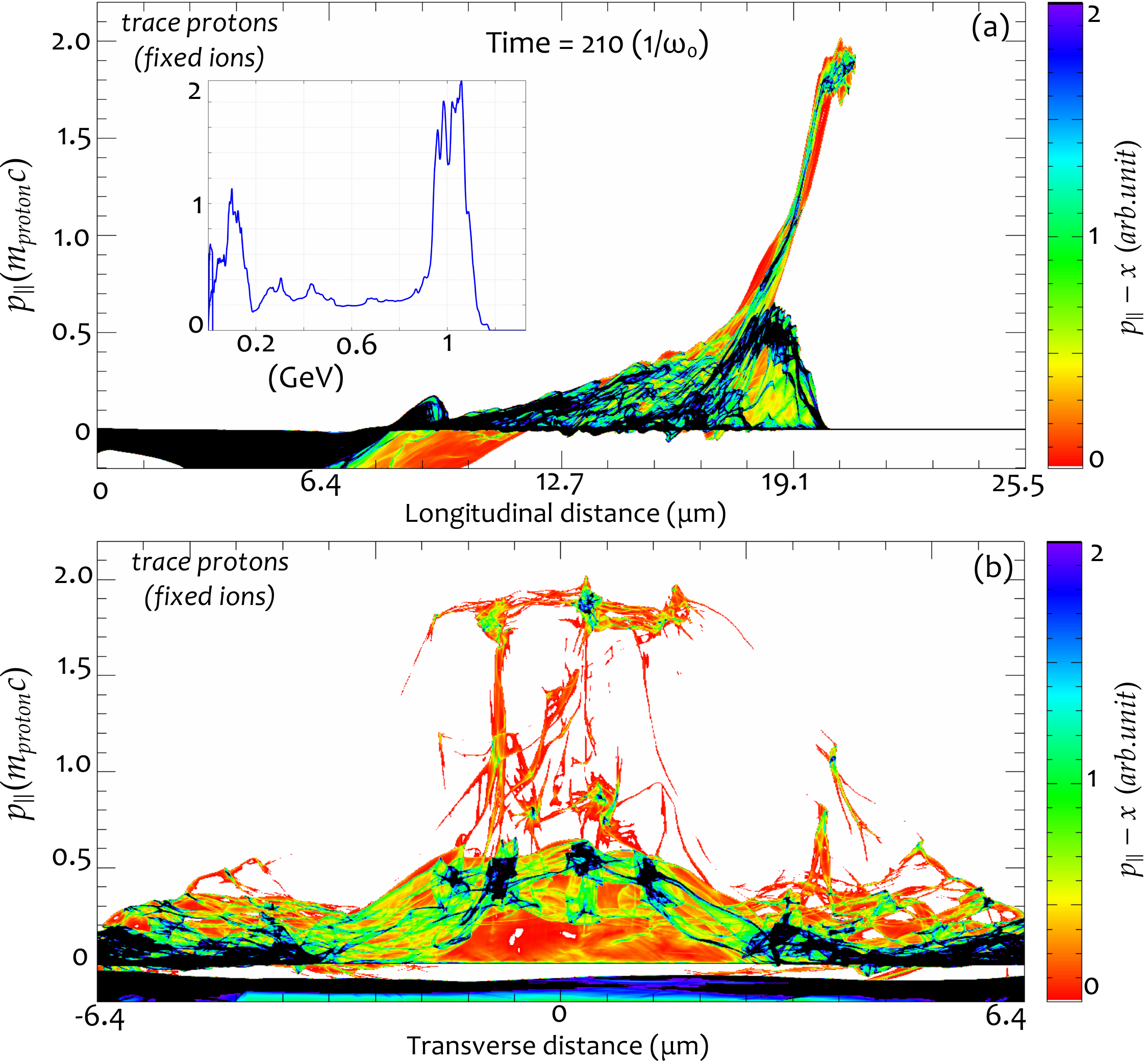}
	\end{center}
\caption{ { \scriptsize {\bf 2-D PIC longitudinal momentum phase-space of trace-proton ($p_{\parallel}$) doped in a heavy background ion plasma for laser-plasma parameters in Fig.\ref{fig:2D-heavy-ion-electrons-high-intensity} }. In (a) the longitudinal momentum phase space of the trace proton species ($0.01n_{crit}$) is along the longitudinal dimension. In (b) it is along the transverse dimension. The protons bunches are accelerated to a kinetic energy of $\simeq 1GeV$ with a mono-energetic spectra as in inset of (a).} }
\label{fig:2D-heavy-ion-proton-high-intensity}
\end{figure}

\begin{figure}[h!]
	\begin{center}
   	\includegraphics[width=3.4in]{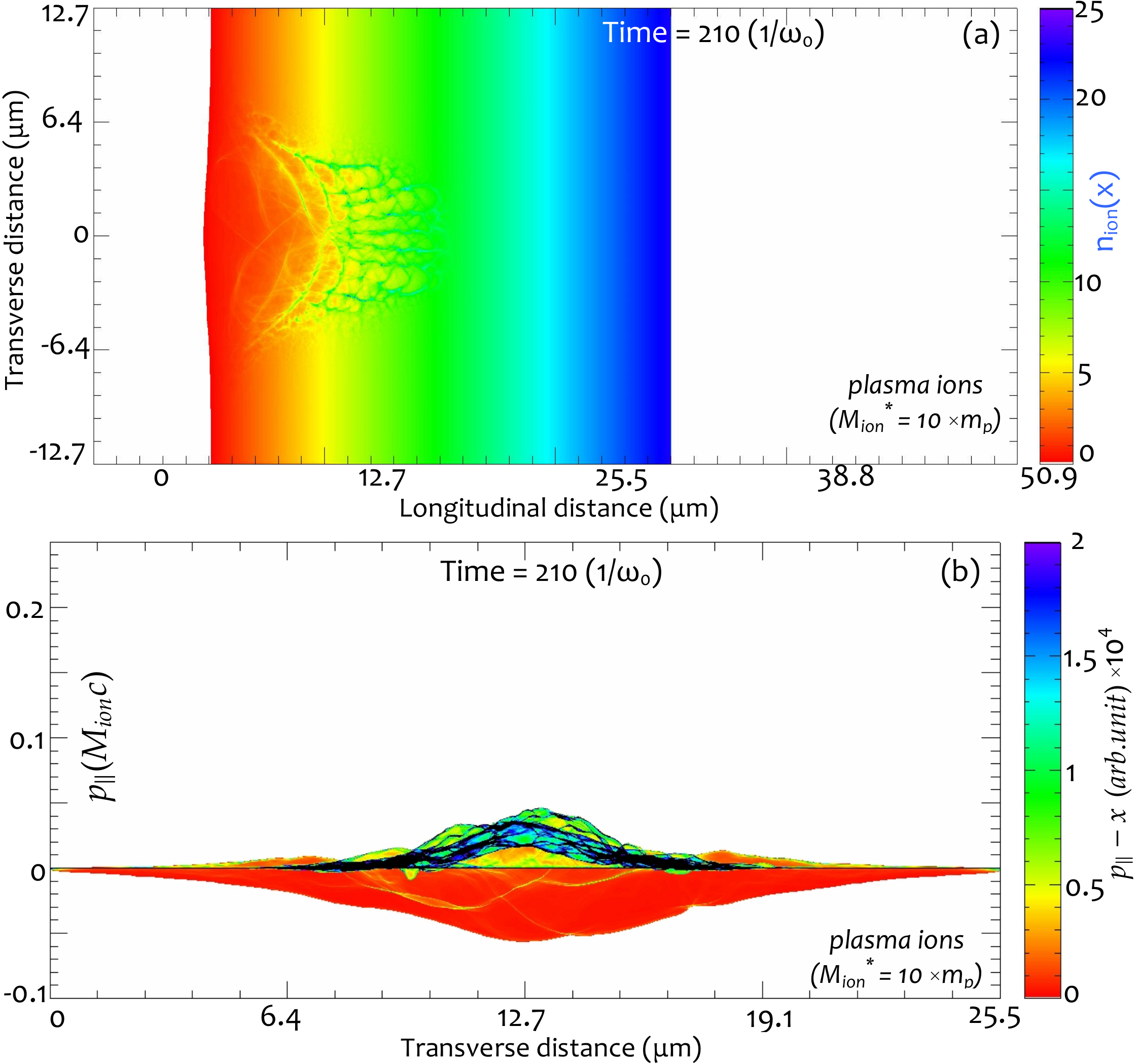}
	\end{center}
\caption{ { \scriptsize {\bf 2-D PIC heavy background ion plasma density and phase-space with plasma with heavy-ion $M^*_{ion} = 10m_p$ for laser-plasma parameters in Fig.\ref{fig:2D-heavy-ion-electrons-high-intensity}.}  The effect of the critical layer motion upon heavy plasma-ions (of realistic $\frac{M_{ion}}{Z}$, $M^*_{ion} = 10m_p$) is negligible. This simulation is a separate computational case to demonstrate the effect of impulse of the electron snowplow potential on heavy-ions. In (a) at $t=210\frac{1}{\omega_0}$ the ion density perturbations in real-space are after the critical layer propagation has nearly stopped (laser intensity stops rising). In phase space (b), ions pick up negligible longitudinal momentum.} }
\label{fig:2D-heavy-ion-ions-high-intensity}
\end{figure}


\begin{figure}[h!]
	\begin{center}
   	\includegraphics[width=3.4in]{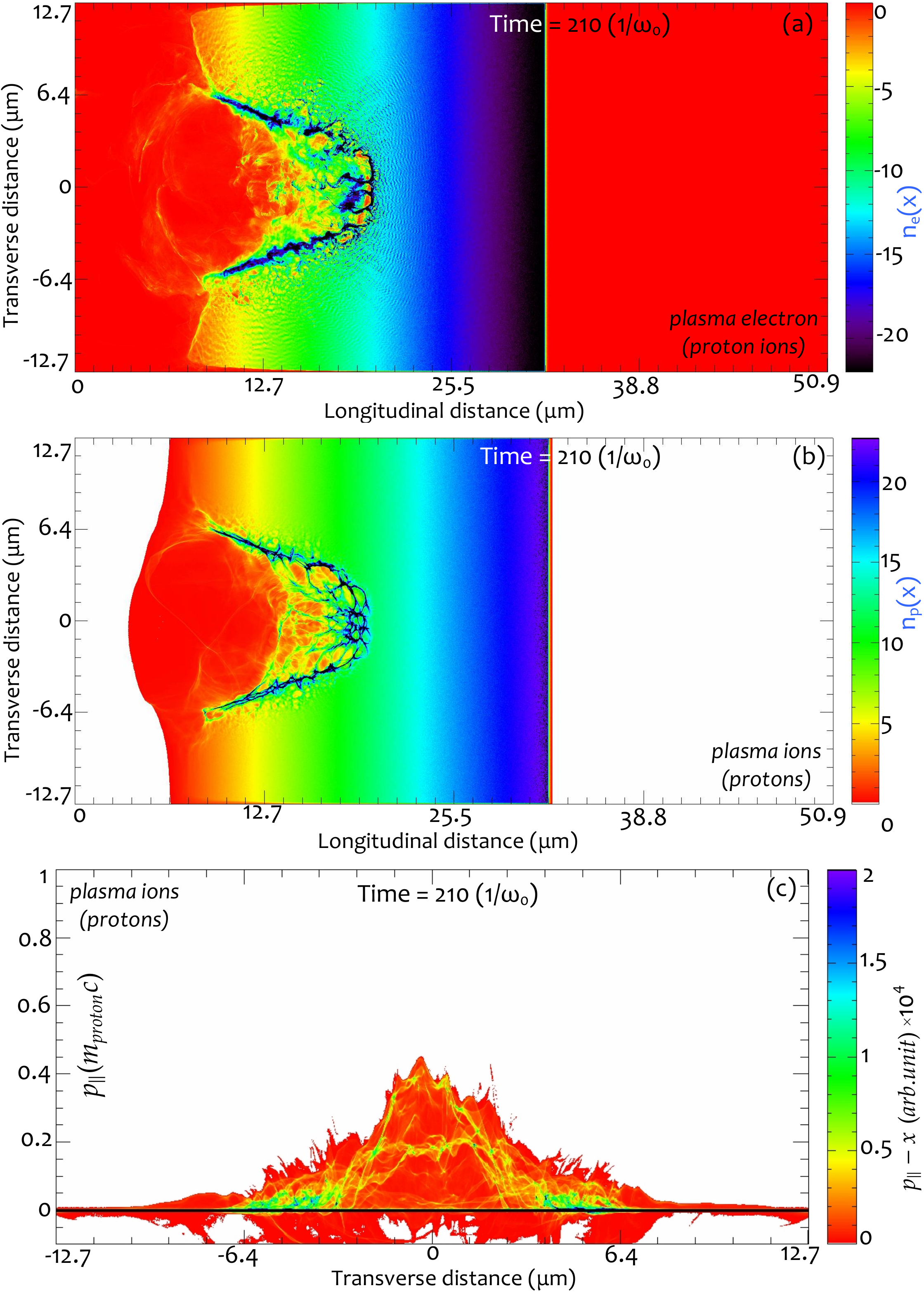}
	\end{center}
\caption{ { \scriptsize {\bf Critical layer motion in a light co-moving ion (proton) plasma density gradient of $\alpha=8.8\frac{c}{\omega_0}$ with laser of $I_0=1.1\times 10^{22}\frac{W}{cm^2}$ ($a_0=72$, $r_0=3.8\mu m$)}. Propagating electron layer is in (a) and the corresponding plasma-ion (proton) layer is in (b) at $t=210\frac{1}{\omega_0}$. In (b) the background plasma-ion density develops perturbations of the order of electron density and co-propagates with the electron layer. The electrons and plasma-ions are nearly evacuated in the region behind the laser front. However, not all the protons are evacuated behind the laser-front therefore hole-boring is incomplete. In (c) the longitudinal momentum phase-space, the plasma-ions at $\simeq 0.4c$ co-propagate with the critical layer at the laser-front.}}
\label{fig:2D-light-ion-high-intensity}
\end{figure}

\section{Conclusion}
\label{sec:conclusion}

The analysis and PIC simulations presented here have established the relativistic plasma critical layer motion in immobile ion plasma by the effect of relativistic electron momentum excitation. It is established that by the phenomenon of relativistically induced transparency the critical layer propagates at a velocity independent of the ion mass. It is validated that in a light co-moving ion plasma the flat-top laser front propagates by exciting a radiation pressure displaced electron layer with high enough potential to drag the plasma-ions. It is also demonstrated that the radiation pressure driven laser front motion occurs in a flat laser intensity profile, not possible in an immobile ion plasma. Due to certain physical effects, the critical layer motion in 2-D does not exactly follow the 1-D model. However, the 1-D model serves to establish the fundamental scaling laws. We have also explored the use of moving acceleration structure co-moving with the critical layer for ion acceleration (LIA) applications. We have shown that in an ubiquitous pre-plasma density gradient of plasma ions, trace protons can be accelerated to relativistic energies. And the difference in the two physical processes when applied to LIA is not incremental.

\section{Acknowledgement}
We acknowledge Prof. T. Katsouleas for his discussions on some topics presented here and for the purchase of tools used. We also acknowledge the access to the 256 node {\it Chanakya} cluster of the Pratt school of engineering at Duke university maintained by Mr. M. Sayed. Work supported by National Science Foundation under NSF-PHY-0936278 and Department of Energy under DE-SC0010012.


} 

\end{document}